\documentclass[prl,showpacs,floatfix,amsmath,amsfonts,twocolumn]{revtex4-1}  
\usepackage{graphics}
\usepackage{epsfig}
\usepackage{color}

\usepackage[normalem]{ulem}

\definecolor{dgreen}{rgb}{0.0, 0.5, 0.0}

\begin{document}

\title{Simple wealth distribution model causing inequality-induced crisis without external shocks}

\author{Henri Benisty}

\affiliation{Laboratoire Charles Fabry, Institut d’Optique Graduate School, CNRS, Univ. Paris Saclay, 2 Ave Augustin Fresnel, 91127 Palaiseau Cedex, France
}
\date{\today}

\begin{abstract} 

We address the issue of the dynamics of wealth accumulation and economic crisis triggered by extreme inequality, attempting to stick to most possibly intrinsic assumptions. Our general framework is that of pure or modified multiplicative processes, basically geometric Brownian motions. In contrast with the usual approach of injecting into such stochastic agent models either specific, idiosyncratic internal nonlinear interaction patterns, or macroscopic disruptive features, we propose a dynamic inequality model where the attainment of a sizable fraction of the total wealth by very few agents induces a crisis regime with strong intermittency, the explicit coupling between the richest and the rest being a mere normalization mechanism, hence with minimal extrinsic assumptions. The model thus harnesses the recognized lack of ergodicity of geometric Brownian motions. It also provides a statistical intuition to the consequences of Thomas Piketty's recent ``$r>g$'' (return rate $>$ growth rate) paradigmatic analysis of very-long-term wealth trends. We suggest that the ``water-divide'' of wealth flow may define effective classes, making an objective entry point to calibrate the model. Consistently, we check that a tax mechanism associated to a few percent relative bias on elementary daily transactions is able to slow or stop the build-up of large wealth. When extreme fluctuations are tamed down to a stationary regime with sizable but steadier inequalities,  it should still offer opportunities to study the dynamics of crisis and the inner effective classes induced through external or internal factors.
\end{abstract}


\maketitle
\section{I. Introduction}
Mathematical models for economy using microscopic agent-based descriptions have attracted a lot of attention in the last few decades ~\cite{Sorne97,Sorne98,Bouch00,Levy,Piane,Chatt04,Chatt07,Patri,Yakov,Ichin,Heinsa,Chakr15}. 
They draw on the rich tools of physics to describe some characteristic observed trends in several complex fields. Notably, various features of the statistical distribution of wealth among individuals or entities (firms, cities, etc.), especially those featuring power-law distributions (Pareto tails or Zipf's law), have been studied within assumptions of simple stochastic ingredients~\cite{Gabai99,Saich08,Saich09,Malev,Perre}.

Furthermore, nowadays, the degree of inequality in wealth distribution as well as its evolution are issues of growing interest. A witness of this worldwide interest, beside the  echo of extreme wealth inequality as yearly reported by Oxfam for instance,  is the success of Thomas Piketty's analysis~\cite{Alvar,Piket}, namely the ``$r > g$'' paradigm (where $r$ is the return rate of capital and $g$ the growth rate of the whole economy): from a ``law'' that is deceivingly simple, historical analysis of long-time series of patrimonial wealth and incomes across centuries suggests that its implications at the multi-decadal scale are possibly very large.

Adverse or beneficial consequences of inequality in agent-based models are mostly thought in terms of some explicit extra variable(s) with threshold or similar procedures that amount, from a physicist's point of view, to nonlinearity. The economics narrative translates this in various ways, within current political biases~\cite{Gilen}: The ``trickle down'' effect suggests that any ``added value'' created by the large means of the affluent shall, sooner or later, diffuse down all social strata of society and incur an overall benefit. Features that can be explicitly considered are for instance the advent of monopolies and how their adverse effects on competition, pricing, innovation, firm creation and death can be tracked. Note also that a majority of wealth distribution studies stick to a static view, even if fine rendering of actual data is sought, including for instance the role of inheritance and bequest strategy~\cite{Benha}.

The generic idea that such nonlinearities or extra parametric bias then induce crisis, and change the growth regime from smooth to moderately or highly intermittent, has been acknowledged generically in the broad wake of John Maynard Keynes, and specifically by the economist Hyman Minsky. More recently, the work by the ``heterodox'' economist Steve Keen could substantiate recent trends in escaping the ``representative agent model'' and its questionable ability to describe crisis mechanisms. In statistical econophysics models, the ``wealth condensation'', that describes the advent of extreme events, extreme inequality in particular~\cite{Bouch00,Yakov,Alvar,Galle}, emerged in the first years of the discipline.

However, the lure of describing accurately Pareto tails or other fat tails epitomizing inequality led to the dynamics of wealth distribution being rarely investigated until a few years ago. The input of econophysics hinged a lot on equilibrium thermodynamics, with its ability to bridge micro to macro and predict emergences such as phase transitions for instance. So, non equilibrium thermodynamics, that is, how wealth distributions adapt to permanently moving landscapes and possibly never relax to a steady state, was rather left aside.

A compounded effect on this state of affair is the fact that the Geometric Brownian Motion (GBM) of ``agent'' ensembles, a tenet of stochastic studies of wealth distributions, leads, when left to evolve freely, to nonstationary distributions: essentially log-normal distributions with drift and broadening width over time. The lack of ergodicity of these ensemble distributions was recently pointed out (i.e. the average wealth of a population at time $t$ does not converge at all, at large times, to the same value as the average of individual asymptotic fates). This is mathematically no more than an issue of noncommuting limits~\cite{PRL}. But the lack of recognition of this issue has plausibly induced several weaknesses in mainstream economics, as recently pointed out~\cite{Gamble, Adamou}, the most striking being the outright rejection of diverging utility functions in models. Studying the heart of the topic, conversely, led to a possibly more intrinsic metric of inequality based on the logarithm~\cite{Gamble, Adamou}. 

Practically, it is of course desirable to confront the ``socially agnostic'' GBM distributions to a description of growing inequalities~\cite{Gabai16}. This can be done by attempting to track the large inequality modulations in the last century through a reallocation mechanism acting as a restoring force~\cite{Berman}. Doing so, it appears that, even without any social bias such as lower education of the poor class or the likes, the best description of the last four decades is one of diverging inequality and negative reallocation. The chiasm is large compared to a conventional ``adiabatic'' picture of a stationary distribution that would evolve close to a local equilibrium with gentle disturbances from economic factors addressed through various ``output gaps'' ~\cite{Berman}. 
	
 The relevance of models in relation with their dynamics rather than their equilibrium distribution is also addressed by Ref.~\cite{Gabai16}, as the study finds that any bare GBM would cause too slow a dynamic, once a plausible stationary calibration of the GBM is done. Only by introducing a nonlinearity related to an inner society structure, can the dynamics be appropriately fitted. The work by Liu and Serota~\cite{Liu} shows that nontrivial nonstationary mathematical distributions can be characterized through the time-constants and correlations of their momenta. So the issue of modelling (and calibrating) both the transverse (population) and longitudinal (time-dependent) wealth distributions is becoming the minimally significant scope to exploit such models and get insight from them.

Along this scope, it seems logical to explore all the dynamics of GBM-based models, and notably involving their most striking feature, related to non ergodicity, the inequality ``condensation'', whereby the wealthiest individuals not only capture a large fraction of the total, but also generate the largest positive or negative fluctuations. For instance, there is no clear intuition whether a few time-constants can be relevant to describe realistics dynamics, or if a wilder evolution, with a ``noisy'' spectrum is the better picture. It is therefore interesting to exploit the background of GBMs to explore such issues, with the view that they can reveal subsets and dynamics that are not an obvious part of the conventional wisdom in the way the science of complex systems is applied to economics.

In this paper, we focus in the spirit exposed above on a model whose intermittence is related to high inequality, but whose nonlinearity is as implicit as possible. We thus avoid to cast a moral stance on microscopic behaviors and their intended modifications. Still, we believe that the way it reveals the dynamics of inequality suggests a vision of the diagnoses to be made in actual economies and of the possible counter-measures to be associated. Such visions could help broadening the much demanded alternative points of view to the so-called conventional wisdom.  

\begin{figure}
   \includegraphics[width=\columnwidth]{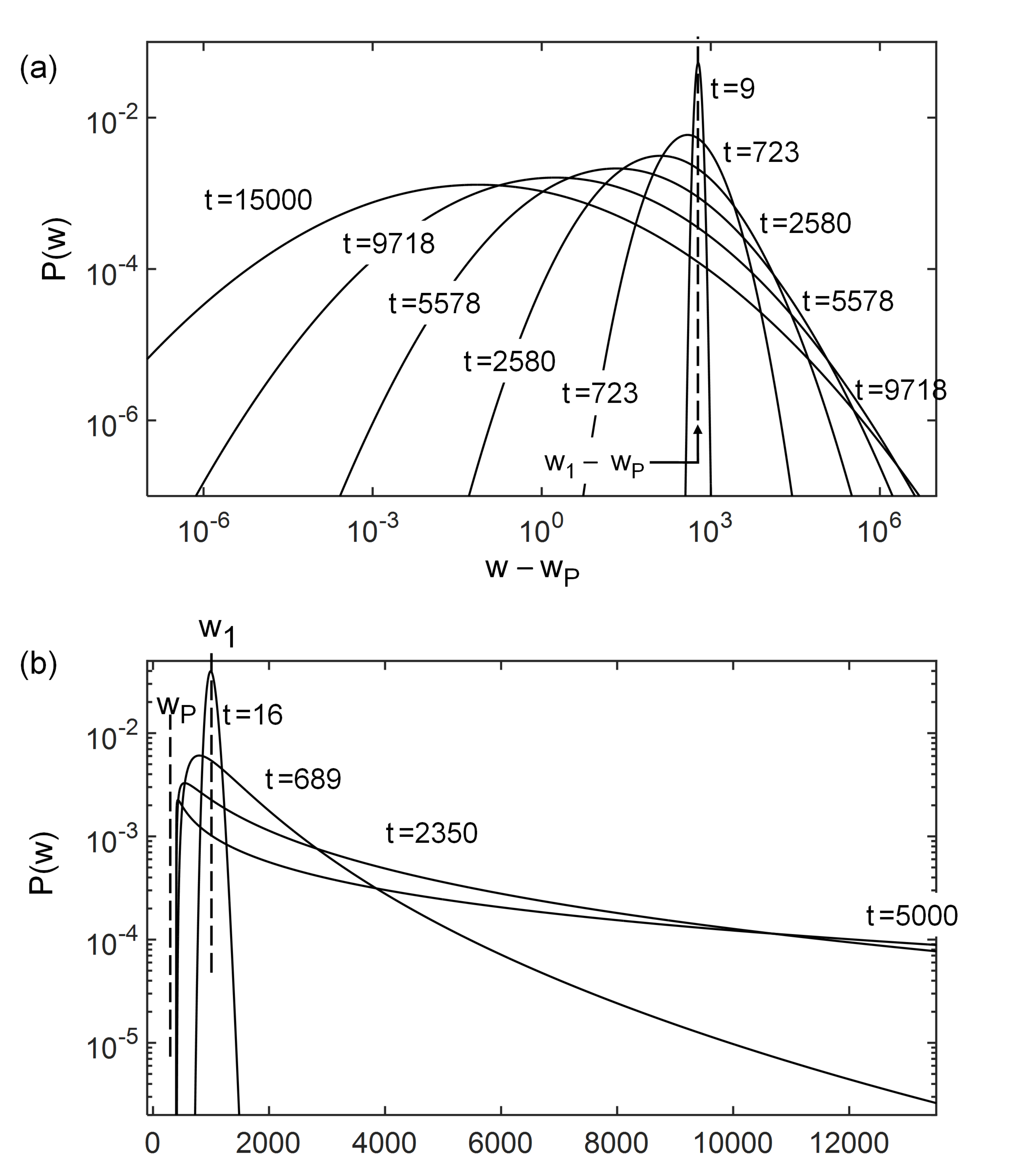}
  	\caption{(Color online) (a) Gaussian distribution of $w-w_p$ at increasing times as indicated (log-scale), with $\beta=0.06$ and initial distribution concentrated at $w_1$; (b) Wealth distribution $P(w)$ on a linear scale.} 
  	\label{fig:one}
  \end{figure}

In a nutshell, we examine a model whereby, according to the oxymoronic say, ``the tail is wagging the dog'', i.e., the presence of a tail of a few very large wealth moves the distribution as a whole. Its discrete nature can be anticipated to blur simplified collective dynamics, e.g. those with simple time constants. The interest of this emphasis is to attract the attention to the general features (in time and in instantaneous distribution) that are likely to link excessive inequality and a situation of uninterrupted crisis reminiscent of the last decades of worldwide economic troubles. Generality is granted here by simplicity, not taking the bias of a ``mean-field'' approach with an average representative agent, but rather outlining the ``space-time'' roles of the extremes~\cite{Chatt16}. Further mapping on various topologies of networks~\cite{Ichin,Vital} could of course help understanding, as well as a cross-analysis with emerging GBM dynamics studies~\cite{Benha,Berman,Gabai16}. Our exploitation will be to identify a ``water divide'' of wealth flow, opening opportunities to view the separated ``basins'' as classes.

The basis of our model is a simple set of $N$ random multiplicative processes that describe the daily fate of agents’ wealth~\cite{Sorne97,Sorne98, Piane, Chatt04, Chatt07, Patri, Yakov, Ichin, Heinsa, Perre, Redner}. Needless to say, multiplicative process are associated to interest rates, but as is usual for GBMs in this context, we do not model anything but ``wealth'', and have no time horizon (i.e. no long term correlation, finite agent life, etc.) in microscopic features. If we denote $w_j(t)$ generic variables at integer times $t$ (days) submitted to a multiplicative process, described by a probability distribution $\Pi(\lambda) d\lambda$ to get  $w_j(t+1)= \lambda w_j(t)$, we have an additive process by considering their logarithm  $x_j =  \log(w_j)$. We initially stick to the balanced case where the average gain expectancy of agents is zero, that is $\int{\lambda\Pi(\lambda)d\lambda}=1$~\cite{Saich09, Malev}. This makes more clear the dynamic role of extremes, as the primary evolution without sizable extremes in the distribution is a gentle unstructured noise, compatible with the impression of a fair game.

In Section II, based on the known log-normal distribution, we develop what happens in this fully independent but yet discretized evolution~\cite{Saich09}: We give a few ``calibration'' hints to justify a rather high daily ``bet'', independent of the Kelly criterion, noting the unsatisfactory status of time constant calibration~\cite{Berman}. We explain what happens when starting from an idealized Dirac-type egalitarian wealth distribution  $P(w)$ (much as in~\cite{Adamou, PRL,Berman} accounts). We use the underlying log-normal distribution of $x_j$, which has a residual nonzero drift because  $\int{\ell\pi(\ell)d\ell<0}$ , where $\ell=\log(\lambda)$  describes the additive process deriving from the multiplicative one~\cite{Sorne97, Sorne98, Saich08, Saich09, Malev}. The evolution of this distribution causes a strong intermittency regime occurring when the wealthiest agents reach a large fraction of total wealth~\cite{Bouch00, Ichin}. The drift component results, at long times, in a global impoverishment of all agents. We assess the role of $\log(N)$ in determining essential time constants. Note that we impose a nonzero (``floor'') lower wealth. Although it has no influence in this Sec.II, it is consistent with the next sections and affects the post-condensation fate.
 
In Section III, we introduce the simple feedback mechanism of normalizing the average wealth to its initial value~\cite{Piane, Gabai99, Saich09, Yakov, Chatt16, Vital}. We show that this results in endless intermittency that affects the whole wealth spectrum. Here, our ``wealth floor'' plays a dynamical role, as it provides a flux even after a large collapse. We confirm the picture that ``the tail is wagging the dog'', i.e., that the wealthiest agent fluctuations are those that impact on the rest, by studying the correlation of ordered wealth series. There are two aspects here: first, we observe how the dynamics can be momentarily much shorter than the $\log(N)$ one. Secondly, we identify a ``water divide'' of wealth flow. There is therefore an effective closure between system size (thus, $N$), system collective dynamics (crisis and intermittency), and this inner divide. 

In Section IV, we deduce how a damping mechanism could act, and summarize the possible meaning and use of the results in Sec.V. If our diagnosis holds, then preventing the build-up of too large entities should be averted. A brief ``stylization'' discussion is made~\cite{Alvar, Piket, Gilen, Galle, Vital}. Our choice in Section IV is to bias the daily ``exchange'', which has zero-change expectation in the non-damped model  [$\int{\lambda\Pi(\lambda)d\lambda}=1$]. We do so by favoring a small gain of the poorest, and a small loss of the richest, by an amount which is very small (0.5-2\%) compared to the daily ``bet''. This choice was inspired by the pricing mechanisms introduced by Aristotle, whereby the price has no absolute underlying reference that the market should ``discover'', but is rather related to the ``social status'' of the agent, as Paul Jorion pointed out from his anthropological studies of various communities~\cite{Jorio}. This view does not really contradict the usual pricing law of supply and demand in a linear regime (a continuum of status), but it allows to extend it to extreme cases, enabling a survival revenue notably, i.e., forms of solidarity that go beyond mere greed and are present in an ``embedded'' view of economy, to use Karl Polanyi's words. We show that small amounts of this bias are effective in suppressing the intermittency and result in a stationary self-replicating wealth distribution. Then the equilibrium distribution can be found as the solution of a tractable eigenvalue problem. The eigenvalue spectrum could also help further dynamical studies. Indeed, a remaining degree of nonstationarity could also be part of the required stylization of human economies. Let us finally underline that we have no substantial consideration for the wealth’s distribution in terms of Pareto tail or power law~\cite{Saich09}, letting this for further work, as obviously there is interest in the topic~\cite{Perre}.

\section{II. The N-agent multiplicative model in the case of mere random walk}

We consider here $N$ agents of wealth $w_j(t)$ at discrete times $t=1,2,\ldots$ To study the wealth distribution intrinsically, we set the average wealth at the start as a constant $w_1$. We want to account for the exchange of wealth and information among agents without any explicit microscopic mechanism. The simplest assumption is that education of agents teaches them to stay on the crest of gain and loss on the average. We then have a zero-sum exchange, with only individual fluctuations~\cite{Sorne98, Piane, Chatt04, Yakov}. We thus model the process as a multiplicative one (often called Gibrat's law) with some simple added features detailed below: at each time (each day) the agent engages a given fraction $\beta$ of his wealth. To make the model easier to grasp in terms of metaphor, we set it up as follows:
\begin{eqnarray}
w_1&=&1000 \label{eqnw1wplambda_1}\label{eqn1w1}\\
w_p&=&400 \label{eqnw1wplambda_2}\label{eqn1wp}\\
w_j(t+1)&=&w_p+\lambda\:(w_j(t)-w_p)
\label{eqnw1wplambda_3}
\end{eqnarray}
where we use for the multiplier $\lambda$ a rectangular distribution uniformly spanning $[1-\beta,1+\beta]$ for simplicity:
\begin{equation}
\Pi(\lambda)=\Pi_0\:\textrm{rect}\left(\frac{\lambda-1}{2\beta}\right) = \frac{1}{2\beta}\:\textrm{rect}\left(\frac{\lambda-1}{2\beta}\right) 
\label{PIrectangular}
\end{equation}
where rect is unity in the interval $\left[-\frac{1}{2},\frac{1}{2}\right]$. Clearly, $\int_0^\infty{\lambda\;\Pi(\lambda)\;d\lambda} =1$ , i.e. there is no ensemble average change in wealth in an individual process. The reader should nevertheless be aware of the nonergodicity of such ensembles~\cite{PRL}. We will see later the time scales at which care is needed, at least when  initial conditions are canonical. Eq.\ref{eqn1wp} introduces $w_p =400$ as a minimum ``floor'' wealth. It is important to set the ratio $\frac{w_p}{w_1}$  to values such as 0.4 here that at least grossly represent developed economies. A simple aspect is that in this way, we will stick to an underlying Brownian motion (a simple Brownian motion for the logarithm of $w$) and preserve a decent value for total wealth at the slumps, without need to introduce a barrier or other nonlinearity at the small wealth end of the distribution. Also, when it later comes to the feedback (Sec.III, Sec.IV), the coupling itself will have noticeable effects because it acts on nonzero wealth for the (many) poorest agents, hence a likely role on dynamics. Thus, while it introduces an unneeded feature in the present section, we retain this poverty ``floor'' here and throughout the paper.

As is well known~\cite{Sorne97,Sorne98, Piane, Yakov, Gabai99, Saich09, Redner, PRL, Adamou}, starting from a given initial state at $t=0$, we have a diffusion+drift process in the $x = \log(w-w_p)$ space. Starting from an initially single-valued distribution $P(w,t=0) \equiv \delta(w-w_1)$, in other words a distribution concentrated at $x_0=\log(w_1-w_p)$, the distribution of the variable $w-w_p$ undergoes two evolutions : it spreads diffusively like a Gaussian in $x$ space, thus taking the form of a parabola in a log-log  representation, and its center also drifts. 

Both effects are determined by the second and first momenta of the distribution of $\ell=\log(\lambda)$, denoted $\pi(\lambda)$, which obeys  $\pi(\ell)\;d\ell =\Pi(\lambda)\;d\lambda$. It is found by standard algebra that in our case of zero-average exchange of Eq.\ref{PIrectangular}, the drift velocity (per unit time in $x$ space) is given by:
\begin{eqnarray}
\nu_\textrm{drift}&=&\frac{1}{2\beta}\left((1+\beta)\log(1+\beta)-(1-\beta)\log(1-\beta)\right)-1 \nonumber\\
&\:\simeq\:&-\frac{{\beta}^2}{6}
\label{Nu_drift}
\end{eqnarray}
where the approximation is valid for small $\beta$. The distribution and its standard deviation $\sigma(t)$ in $x$-space take the form:
\begin{eqnarray}
\hat{P}(x,t)\;&=&\;\frac{1}{2\pi\sqrt{t}}\;\exp\left(-\frac{(x-(x_0+\nu_\textrm{drift}t))^2}{2\sigma(t)^2}\right)
\label{Pxt}\\
\sigma(t)&=&2\beta\;\sqrt{\frac{t}{4\pi}}
\label{sigma_t}
\end{eqnarray}
as depicted in Fig.\ref{fig:one}a in log scale of $w$ and Fig.\ref{fig:one}b in linear scale of $w$.
The fact that there is a drift for a zero-exchange distribution is one of the several not-so-intuitive aspects of GBMs (see the example of Fig.2 in~\cite{Gamble} pinpointing the less-intuitive aspect of multiplicative vs. additive processes, as appears from the authors account of a large swath of scientific history through two Bernoulli family members and Laplace).

Apart from the two wealth ingredients $w_1$ and $w_p$, we have to discuss a realistic value for $\beta$. If we had to discuss about agents faced to optimizing their utility when risking their wealth in some bet, we could recourse to the Kelly criterion, for instance. But our scope is distinctly different: we rather want to calibrate the randomness of the economy as a whole (and not as a stationary distribution, but rather vs. its dynamics), and thus in our spirit, the choice of $\beta$ rather has to be dictated by the spread of fate for a bunch of agents of any given initial wealth: the relative spread $\frac{\Delta(w-w_p)}{w-w_p}$ is independent of initial $w$ in a multiplicative process. Also, while a multiplicative process resembles an interest rate, we should not follow this analogy, as an interest rate is a drift, not a spread (see~\cite{Gabai16} and ~\cite{Berman} for calibration issues with GBMs). By considering the width of the distribution at a typical time of economies, $t=1000$ days, about 3 years (and less than typical periodicity of economic cycles, say 8-40 years), we chose to home in on relatively large variations~\cite{Bouch00, Patri}: Using $\beta=0.06$, we have $\nu_\textrm{drift}=-0.0036$  (per day) and for instance $\sigma(t=1000\:\textrm{days}=1.07$). This seems large, but the corresponding characteristic factor $e^\sigma=2.92$ should be applied to $(w-w_p)$, whose most frequent or median value is expected to be closer to $w_p$ than to $w_1$. So we describe, over a duration of 1000 days,  a spread from, say $w=600$ ($w=w_p+200$) to the 1-standard-deviation interval $[400+200e^{-\sigma},\:400+200e^\sigma]\simeq[469,\:983]$. See the discussion section for further comments.

\begin{figure}
   \includegraphics[width=\columnwidth]{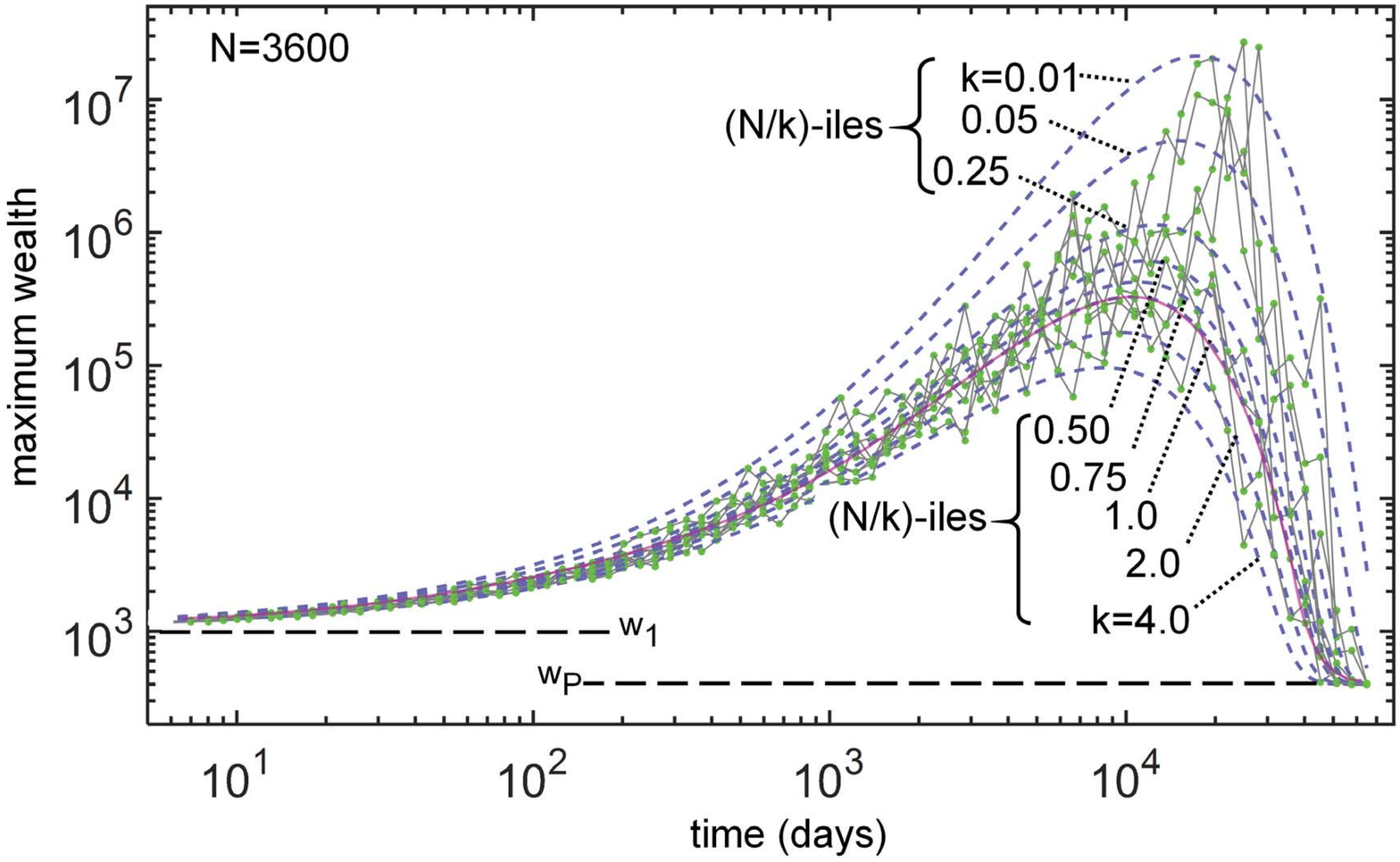}
  	\caption{(Color online) Log-log plot of N/k-iles $x_{N/k}(t)$ (dashed lines) as a function of time for $N=3\,600,\;\beta=0.06$, as defined by Eq.\ref{xnkt}, i.e. the point in the Gaussian distribution such that the partial probability above $x_{N/k}(t)$ is $(k/N)$. The curve $k = 1$ is superimposed as a solid dark magenta line. The largest of N wealth lies, around and above $x_N(t)$, plotted with an added solid line. The largest wealth from several simulations are shown at selected time points (green dots), the fraction of points above $x_{N/k}(t)$ at a given time t showing the expected rarefaction trend with decreasing $k$.} 
  	\label{fig:two}
\end{figure}
	
We now turn our attention to the fate of the whole set of $N$ agents. We pinpoint the role of the wealthiest agents in determining the overall fate of the ensemble~\cite{Sorne97, Bouch00, Piane, Saich09, Alvar, Vital}. Since we have an analytical form of the wealth distribution, we can deduce the statistics of the wealthiest agents at time $t$~\cite{Redner}. We do not embark on this exercise rigorously~\cite{Saich09}, though, we only remark in passing that extremes are key to non-ergodicity demonstrations (Eq. (7) in Ref.~\cite{PRL}), but rarely explicit. We can get enough indication of the location of the maxima by a simpler use of the Gaussian normal distribution ~\cite{Redner}: if we have $N$ realizations of a Gaussian variable, we have a good approximation of the statistics of the largest by slicing the Gaussian into $N$ slices of even weight (see work on electron relaxation bottleneck in quantum boxes~\cite{Benis93, Benis95} for a similar use of a Poissonian statistics). The $N$-th slice is an acceptable approximation of the distribution of the largest wealth, in spite of its abrupt cut-off, avoiding the more tedious math of the theory of extrema~\cite{Saich09, Redner}. What is interesting for us is the ability to use the complementary error function $\textrm{erfc}(x)$ and its inverse $\textrm{erfc}^{-1}$  to deal with the main aspect of such statistics. This is simpler to grasp, for those less familiar with extremal laws, than using the exact Gumbel law, i.e. expressing the cumulant $U(x)$ of the largest value distribution in a form often denoted $U(x)=\exp(-\exp(-z(x,N)))$, with $z(x,N) = [x-a(N)]/b(N)$ with analytical expressions of $a(N)$ and $b(N)$: the typical error by taking this naive mean instead of the exact one is $\sim 0.2$ standard deviation, thus an $(e^{0.2}-1)\sim 20\%$ error.

With just a little more generality, we look for the edge $x_{N/k}$ of what we can term as the $N/k$-th slice, with $k=1$ for the largest, $k=2$ for the two largest, etc. in the spirit of quantiles~\cite{Saich09, Alvar, Piket}. And we also allow ourselves to use a fractional $k$, e.g. $k=0.25$ in $N/k$ in order to target a range reached by the largest variable only in 25\% of the statistical events. We shall call these approximate quantile boundaries those of the $N/k$-iles, generalizing on deciles or centiles, notably in the plot of Fig.\ref{fig:two} below. Specifically, using standard Gaussian statistics, we identify the moving edge $x_{N/k}$ of the $N/k$-th slice accounting for the $N/k$-iles statistics such that 
  				
\begin{eqnarray}
\int_{x_{N/k}(t)}^\infty{\hat{P}(x,t)\;dx}\;=\;\frac{k}{N}\;\;\;\;\;\;\;\;\;\;\;\;\;\;\;\;\;\;\;\;\;\;\;\;\; \label{quantileNDef}\\
x_{N/k}(t)=(x_0+\nu_\textrm{drift}t)+\sqrt{2}\,\sigma(t)\;\textrm{erfc}^{-1}\left(\frac{k}{N}\right) \label{xnkt}
\end{eqnarray}

Note that the second formula takes into account the center drift. 
If we run a simulation of $N$ agents during a long enough time, we expect $x_{N/k}(t)$ to first feel the influence of the diffusion and the $\textrm{erfc}^{-1}$ function. In a log-log scale, the largest element ($k=1$) is the upper envelope of the set of the $N\;\;\;x_j(t) = \log\left(w_j(t)\right)$ traces. If we neglect drift, at short times, we have $x_{N/k}(t)-x_0 \propto \sqrt{t}=\exp(\log(t/2))$, so we start with a set of rising exponentials with just different coefficients as a function of $k$, namely the coefficients $\textrm{erfc}^{-1}(k/N)$  whose trend against $k$ is logarithmic. This can be seen on the left of Fig.\ref{fig:two}, for $N=3\,600$, where a set of several $k$ values is represented, with $k=1$, i.e. the $N$-ile of the Gaussian, shown as a magenta solid curve superimposed  over the set of dashed-lines for other $k$ values. 

Now, at long times, since we have a negative drift, as illustrated by Fig.\ref{fig:one}a, even though the standard deviation grows, the average locations of the $N/k$-iles must all, sooner or later, shift left to $x\rightarrow -\infty$  . More precisely, the smaller $k$ the later the trend of increasing  $x_{N/k}(t)$ (i.e. diffusion) is reversed by drift to a decreasing one~\cite{Saich09}. This is the essence of the mechanism allowing the breakdown of GBM ergodicity as noted in~\cite{PRL}. In other words, it demands too small a fraction of the sample (much less than $N^{-1}$, thus less than one agent) to get a chance of realizing high values in the tail, even though such values carry a major contribution to the (unbound) expectation value in a continuum view.

\begin{figure}
   \includegraphics[width=\columnwidth]{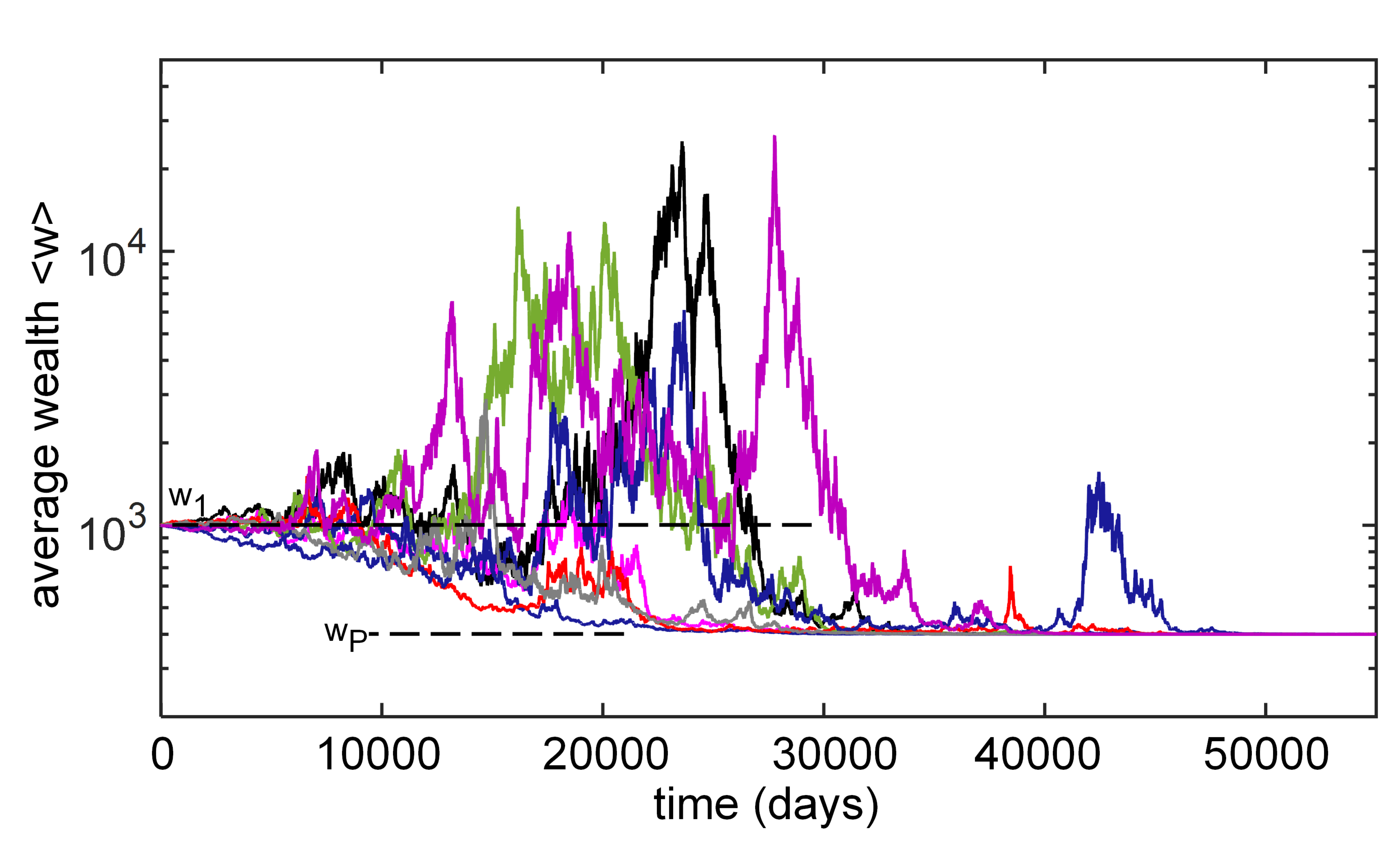}
  	\caption{(Color online) Average wealth of eight simulations (solid lines of different colors) with the same parameters $N=3\,600,\;\beta=0.06$. The initial Brownian-like motion around $w_1$ (level indicated by the right-side dashed line) is gradually suffering larger and larger fluctuation, essentially associated to the large $x_N(t)$ at intermediate times ($t \sim 15\,000 \-- 30\,000$). At larger times, the drift eventually dominates and the average wealth is stuck to the poverty level $w_p$ (indicated by the left-side dashed line).} 
  	\label{fig:three}
\end{figure}
We have added on Fig.\ref{fig:two} the plot of extremal values at selected logarithmically spaced times drawn from a set of eight numerical simulations up to tmax = 55\,000 ``days'' (about 150 ``years''), using the resident random MATLAB generator. Let us comment for instance the outsiders which are the highest points situated around the  ``N/0.01-ile'' curve. Since we chose to sample about 75 points per curve, hence 600 points for 8 curves, the hundred-times rarefaction vs. $k = 1$ entails, probabilistically, a number of points around the $N/0.01$-iles of the order of 6. If we look only around the maximum of the data, the trend-reversal region of the curves around $t=20\,000\;\pm\;15\,000$, where these outsider points are more clearly seen than at earlier times, we are concerned with a subset of about 160 points, and we find around 2 to 4 points in this subset instead of the 1.6 expectation, a reasonable amount for a random draw of this kind (factoring also our approximate extremal law). 

The above exercise is useful to grasp how the Gaussian statistics can be sampled along a long time series: we may, at some times, and provided that we are not restrained by correlation (hence at the lower limit, not at the scale of two adjacent times with $w(t+1)$ and $w(t)$ separated by less than a factor $\beta$)~\cite{Liu}, reach values much higher than $x_N(t)$ in a given simulation.

Let us now focus on the global wealth. Although the statistical average of a single operation is zero, actual operations have some nonzero average. This  emphasizes the role of discretization~\cite{Yakov, Saich09}(again, ergodicity breakdown is the overarching issue~\cite{PRL,Adamou}). At the start, with all $w$'s of the same order $\sim w_1$, fluctuations of this origin cancel out reasonably well: as $N^{-1/2}$  at a given time. Further along the time series, they pull also randomly up or down. So for some time after the start, the average wealth gently performs a Brownian random walk around  $w_1$ (and the total wealth around $Nw_1$), as seen on the very left side of Fig.\ref{fig:three} on a linear time scale.

But, as time goes and the largest agents wealth samples values around $x_{N/k}(t)$  for $k=1$ easily, and further at even smaller $k$ values, large fluctuations are introduced on the total, and thus on the average wealth. This is the ``tail of the dog'', but here the tail is not ``wagging the dog'' forever, as actually there is independence of the $N$ agents wealth, so the stronger fluctuations of the average wealth only reflect the inescapable maximum regions of the curves of Fig.\ref{fig:two}, when the diffusion to large wealth is compensated by the slow drift in the $x$-space (also an onset of apparent ergodicity breakdown). The $\beta^2$ scaling of the drift velocity shows how discretization, and thus $\beta$, is a significant (but not critical) parameter. 
 
We can deduce the order of magnitude of most quantities as a function of $N$ and $\beta$, the sole relevant parameters at this stage, using the approximate drift and the approximate extremal law: 
\begin{eqnarray}
x_{N/k}(t)&\simeq& \left( x_0-\frac{\beta^2}{6}t\right)+\sqrt{\frac{2}{\pi}}\beta\sqrt{t}\;\textrm{erfc}^{-1}\left(\frac{k}{N}\right) \\
t_{N/k}^{\max} &\simeq& \frac{18}{\pi\beta^2}\left(\textrm{erfc}^{-1}\left(\frac{k}{N}\right) \right)^2\\
x_{N/k}^{\max} &\simeq& x_0+\frac{3}{\pi}\;\left(\textrm{erfc}^{-1}\left(\frac{k}{N}\right) \right)^2
\label{xnkmax}
\end{eqnarray}

where the two last lines point the maximum of the first line. For instance in our case $N=3\,600$, we obtain for the $N$-ile position: $t_{N}^{\max} \simeq 3\,600$  and $x_N^{\max}-x_0 \simeq 6.31=2.74 \: \log(10)$, meaning that the edge of the richest $N$-ile slice is around  $(w-w_p) \sim 10^{2.74\,}(w_1-w_p) = 550 (w_1-w_p) \simeq 3.30\;10^5$ (the crude approximation  $\textrm{erfc}^{-1}(u)\simeq[-\log(u)]^{1/2}$ is thus too coarse).

Hence, we can also understand that the typical time frame of the large intermittency window can be assigned a practical interval such as $[t_{N}^{\max},\;\log(100)\,t_{N}^{\max}]\simeq[10\,520,48\,530]$, with an upper boundary taken here so as to be exceeded only in one out of 100 cases: the statistical character of this upper boundary is apparent through the presence of a smaller but clear isolated intermittency peak at $t \sim 42\,000 \sim \log(50)t_{N}^{\max}$ in one of our eight simulations. The typical time scale of the intermittencies is more difficult to provide~\cite{Patri, Ichin, Heinsa}, but it is logical that it appears as only a fraction of $t_{N}^{\max}$ because it takes less time than this for extreme fluctuations to enter and leave the extreme domain ($t_{N}^{\max}$ is the time to go from average wealth to extreme wealth for the fastest of $N$ elements). From the point of view of  realism~\cite{Bouch00}, since $t_{N}^{\max}$ is about 30 years, we have here a confirmation that our $\beta=0.06$ daily value is not that large: this time scale of 30 years is not incongruous with that of actual major crisis in capitalist economies.

Last, the typical large deviations of the average are on the order of a few times the naive quantity $(w_1-w_p)\exp(x_N^{\max})/N$ found when counting the role of a single wealthy agent as causing the fluctuation in the mean: this quantity associated to $k=1$ is small, $(w_1-w_p)\times 550/3\,600=91.7$. But we have enough time in one run, given the relatively flat situation around $t_N^{\max}$, to sample rarefied maxima with smaller $k$, typically $k \sim 0.25$ in one run. Then the above quantity becomes  $(w_1-w_p)\exp(x_{N/4}^{\max})/N$, which is about 2\,000. On few (statistically on one) of our eight runs, we can of course experience eight times scarcer cases ($k=\frac{1}{32}$) reaching a maximum average at  $\sim 12\,000$.

We now comment the $\beta=0.06$ connection with $N$. If one takes for $N$ the population of a large city, $N \sim 10^7$, then, since $3\,600^2 \sim 1.2\,10^7$ , we get at given $\beta$ essentially a doubling of $x_N^{\max}$ and $t_N^{\max}$  since $\left[\textrm{erfc}^{-1}\left(\frac{1}{N}\right)\right]^2 \sim \log(N)$. So if we wish to retain the same characteristic time, a couple of decades, we have to modulate $\beta=0.06$ by a factor $\sqrt{2}$ (and going to the world population $N \sim 10^{10}$ would demand a factor of 2~\cite{Bouch00, Chatt04, Patri}).

Having explored an ensemble of $N$ perfectly non interacting agents, we next implement a mechanism for the tail to be ``wagging the dog''. Overall, the dynamics of the global wealth of our simple non-interacting set of $N$ GBM epitomizes how delicate it is to describe the region of largest manifestation of ergodicity breakdown. We conjecture that clarifying its dynamics would be helpful for the understanding of related models.

\section{III. The N-agent multiplicative model with reset average}

We now consider a feedback mechanism that consists in resetting the average at its initial $w_1$ value at all times. This will therefore introduce correlations~\cite{Liu} and will imprint the intermittent dynamics of the total wealth onto the whole distribution, paralleling the way emerging social structures affect all corners of society. The above model of independent agents with its indefinite downward drift that accumulates all agents wealth to the poverty ``floor''cannot inspire even a stylized description of economic reality. We do not want to affect the multiplicative aspect (the GBM), however. Technically, we simply impose, for all agents $j$ :
\begin{equation}
w_j(t+1)=\lambda(j,t)\,w_j(t)\times\frac{\sum_{m=1}^N\lambda(m,t)\,w_m(t)}{N\,w_1}
\label{Resetw}
\end{equation}

where we made explicit the random variable draw for the j-th agent at time t, following the distribution law of Eq.\ref{eqnw1wplambda_3}.

Most usual discussions of ``normalization'' in agent-based simulations are about the growth rate. They are also invoked in several works on random Brownian motion and thus GBM, but we could not find the consequences that we find here~\cite{Sorne97,Bouch00,Piane,Yakov,Gabai99,Saich09}. We shall discuss in Sec.V the socio-economic meaning of this choice.

\begin{figure}
   \includegraphics[width=\columnwidth]{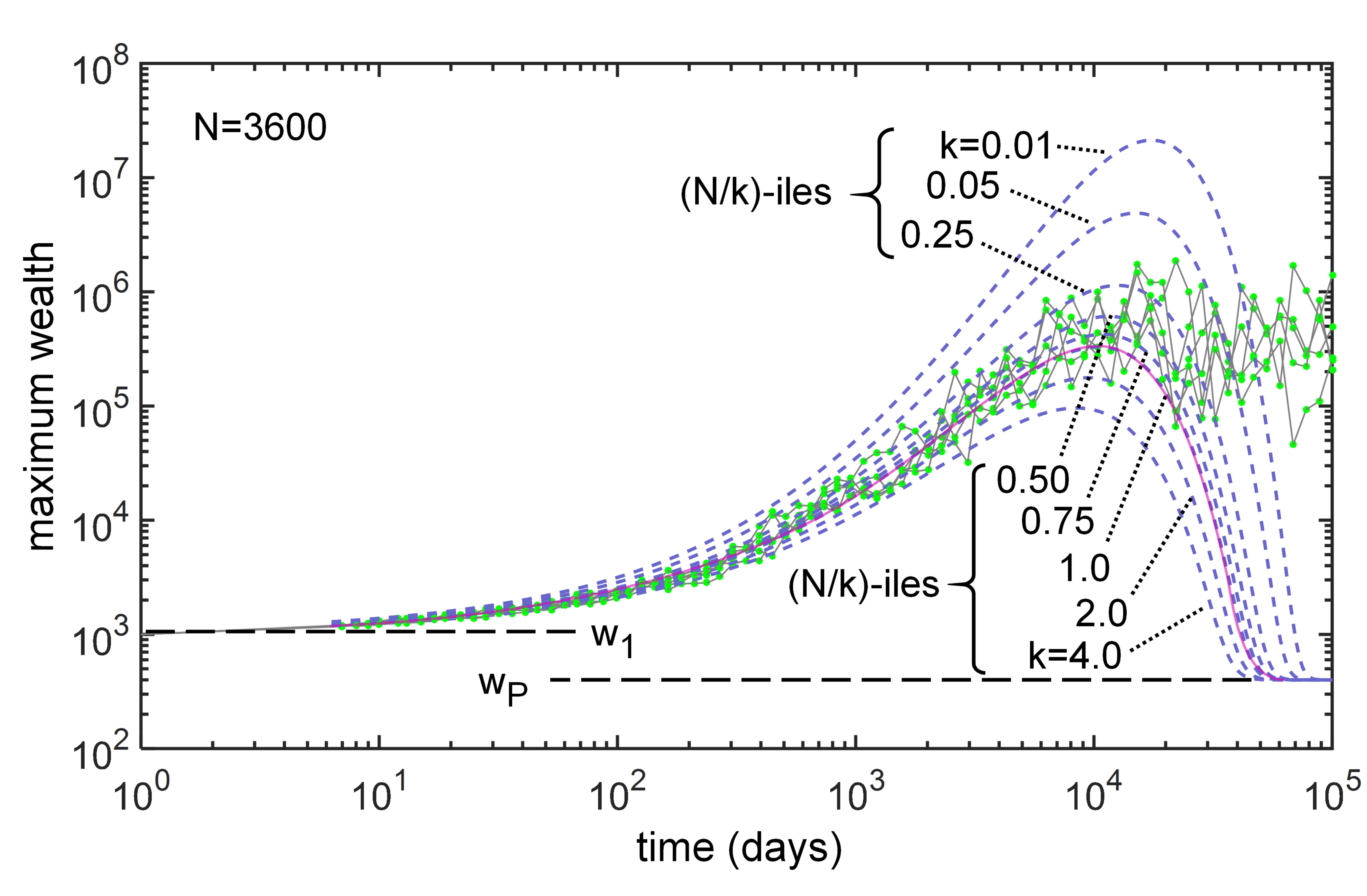}
  	\caption{(Color online): Same log-log plot of maximum wealth vs. time as Fig.\ref{fig:two}, in the case of reset average, i.e. forced normalization of the total wealth to $N\,w_1$. After reaching the point of maximum wealth expectation without reset, the permanent regime retains a fluctuation pattern similar to that occurring around the maximum.} 
  	\label{fig:four}
\end{figure}

\begin{figure}
\includegraphics[width=\columnwidth]{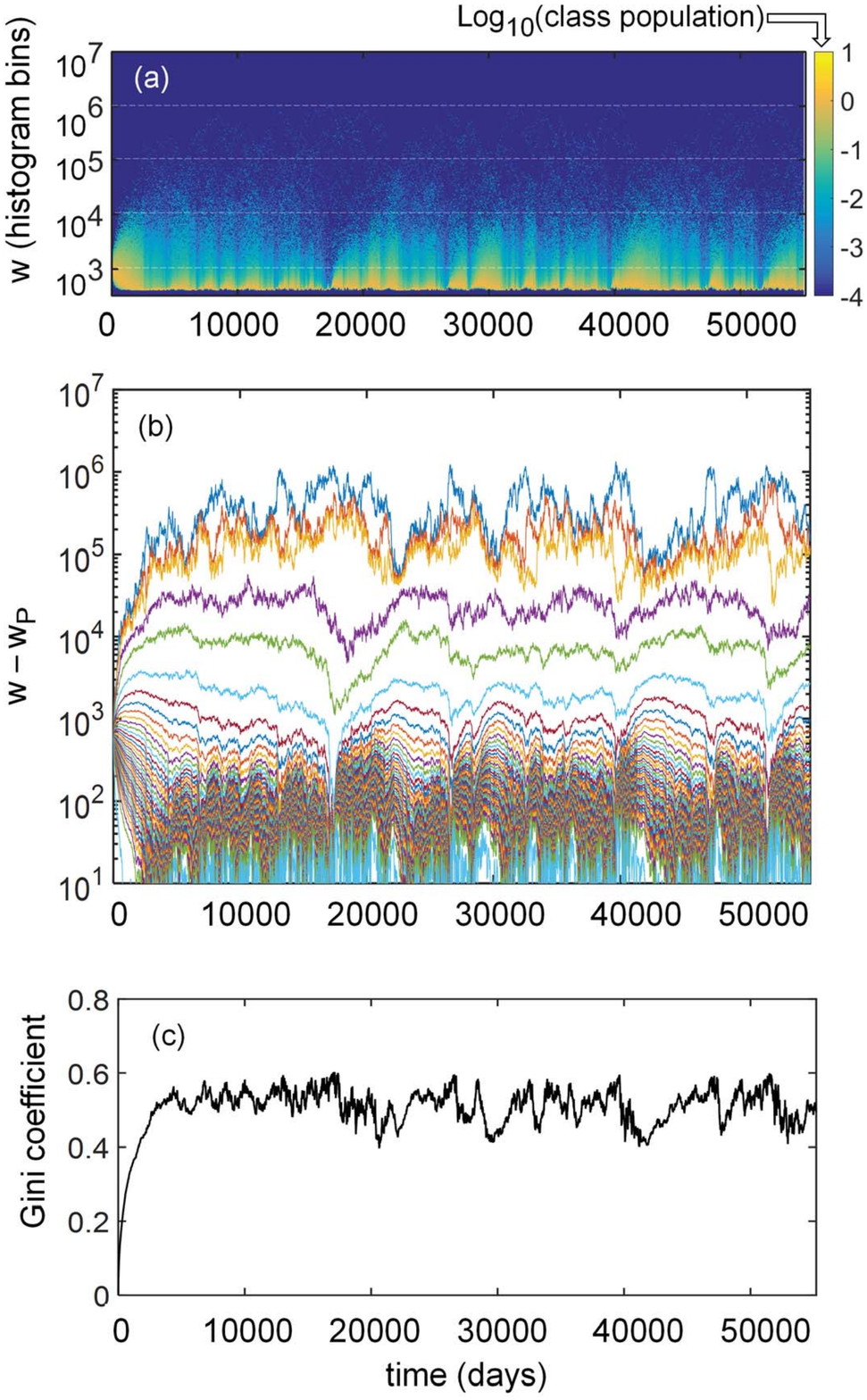}
\caption{(Color online): Illustration of a typical sample of wealth $w_N(t)$ under the reset-to-average assumption. (a) Histogram of wealth using a few hundred log-spaced bins, showing intermittency at multiple time scales; (b) the sorted trajectories of $w_j-w_p$ for selected $w_j$'s. Note the correlation among the majority of small wealth, but note that there is rather an anticorrelation among this vast majority and the top three wealths; (c) Gini coefficient of the distribution vs. time, with  large random fluctuations that still bear the signatures of intermittency of the few wealthiest agents.} 
\label{fig:five}
\end{figure}

In our model, it is most instructive to visualize the fate of the extreme wealth and that of the whole distribution under this new assumption, as we propose through Fig.\ref{fig:four} and Fig.\ref{fig:five}(a-c) respectively. Of course the word ``fate'' means here all the coupled dynamics of society and inequality, with its distribution of events, correlations, and characteristic times.  

In Fig.\ref{fig:four}, we see that under the new assumption of mean normalization, once large wealth are obtained, the downward drift is canceled: the maximal wealth remain around the established level $x_N^{\max}$, and display large fluctuations. We thus operate permanently at the brink of ergodicity. We give a microscopic look into the distribution of wealth thanks to Fig.\ref{fig:five}(a), a color map of histograms fabricated at linearly spaced times $t$ (spacing $\Delta t \sim ~30$ days), by integrating over $\Delta t $.

We can now see that there are collective collapses of the wealth distribution core, down to values close to $w_p$, as soon as there is a chance that the largest value reaches $x_N^{\max}$, say from $t\sim 3\,000$ days on (the analytical value $x_N^{\max}$ can be reached for the first time at a moderate fraction of $t_N^{\max}$, not surprisingly from the general above analysis of Sec.II).

Also striking is the fact that these collapses are followed by revivals, some of them as developed as the start sequence (where we remind that all $w_j$ start from $w_1$), with an overall intermittency pattern. Since we have seen above that there would be occurrences of large average wealth deviations, and that they are due to very few individuals down to a single one, we infer that the same mechanism works here: once an agent is becoming the wealthiest in a steady enough way (that perturbs only marginally the distribution), it suffices that this wealthy agent undergoes larger fluctuations to induce an overall fluctuation of the masses (see the treatment of firms death in Ref.~\cite{Saich09} with extremal law statistics: it provides a resembling, but not directly comparable pattern). That is what we picture as ``the tail wagging the dog''.

In terms of current analysis of GBM-based economics models, we should be looking at time-constant distributions of different classes of agents, and at their correlation. This would be a highly rewarding analysis if the resulting dynamics can be correlated to the available economic data through not only the trends of inequality, but also economic, sectoral and territorial discrepancies. The fact that we can see a ``tail'' and a ``dog'' also points the possibility to distinguish social classes and their line of divide, as will be developed briefly.

The detail of the wealth fate can be perceived in Fig.5(b), where we plot on a log scale a selection of the sorted temporal profiles of $w_j(t)-w_p$ . We clearly see that the troughs apparent in the tail of the poor agents correspond to the aftermath of a peak of the very few wealthiest agents. In other words, in a zero-sum exchange game due to the fixed mean wealth, any of the larger-than-average fluctuations of the wealthiest is felt by essentially all agents, and can be felt as a big shock. 

Let us present our distributions of wealth as a preferred economic indicator. Although fundamental ones have recently been proposed in relations with GBMs ~\cite{Adamou,Gamble}, we present in Fig.\ref{fig:five}(c) the well-known Gini coefficient~\cite{Chatt16, Chatt16b}. Its variations are clearly triggered by the few wealthiest agents. The curve shapes are not identical, but the major  peaks, troughs and shoulders are clearly correlated. These curves contain both (i) the dynamics of inequality in terms of distribution of time constants, insofar as a picture of subsets with a reasonable stationary distribution of time constants applies, (ii) an indication of the amount of correlation within agents, as they account for the amplitude of the fluctuation. 

As for the possibility to define subsets, since we have seen that the largest fluctuations are the leading events, we try below to define two dynamical ``classes'', separated by a ``water divide'' line of wealth flow. Such a picture may provide an account of the actual GBMs mechanisms and invite resonances for the stylization step between model and economic reality. 

\begin{figure}
   \includegraphics[width=\columnwidth]{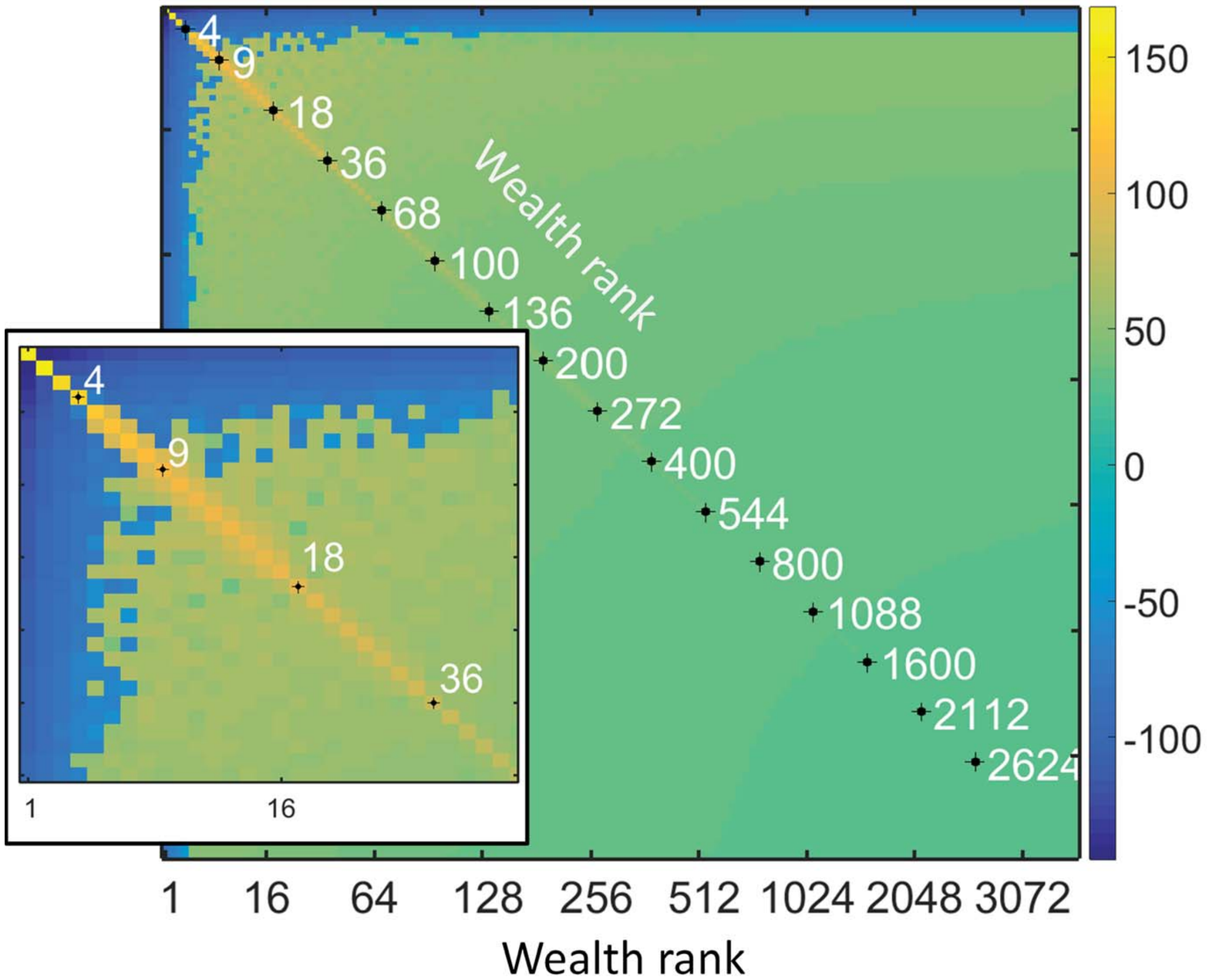}
  	\caption{(Color online): Correlation analysis on ranked wealth. See text and Eq.\ref{logtypeindic} on the particular correlation-based indicator $C_{i,j}$ used here. We represent as a color a quantity measuring the fluxes and their signs, between the affluent and the poorest of the agents, with a step-wise logarithmic sampling of the $3\,600 \times 3\,600$ matrix. Note that the anticorrelation is neatly defined, and that it clearly stems in this graph from gains of the few richest. The inset on the bottom left is a zoom on the $\sim 40$ wealthiest agents.} 
  	\label{fig:six}
\end{figure}

Specifically, we elaborate a color map of a matrix describing the flux pattern between agent pairs $(i,j)$, as is done in Fig.\ref{fig:six}. Mathematically, we work on the sorted series of wealth: we take first the product of the time series of the sorted wealth derivatives,  $y_j=dw_j/dt$ , and we then fabricate a log-type indicator of the absolute value, but we keep track of the sign of the product to distinguish between gain and losses:

\begin{eqnarray}
A_{i,j}&=&\sum_{t=1}^{t=t_{\max}}\left[w_i^{(s)}(t+1)-w_i^{(s)}(t)\right]\,\left[w_j^{(s)}(t+1)-w_j^{(s)}(t)\right]\nonumber\\
&\equiv&  \int_{t=0}^{t=t_{\max}}\left[\frac{dw_i^{(s)}}{dt}\right]\,\left[\frac{dw_j^{(s)}}{dt}\right]\,dt\nonumber \\
C_{i,j}&=&\textrm{sign}(A_{i,j})\,\left[\log(A_{i,j})\right]^2,
\label{logtypeindic}
\end{eqnarray}

where the superscript $(s)$ denotes the sorted ensemble (dynamical sorting at all time $t$, so it scrambles the actual agents throughout the simulation time).

In Fig.\ref{fig:six}, we clearly see a negative correlation between the first four agents ($i=1$ to 4) and all agents beyond  $j \sim 200$. The zoom on the low-$(i,j)$-corner at the top-left of this matrix, shows that till values of the indices $(i,j)   \sim 12$, the correlation is still not clear cut, although the trend toward positive correlation increases for smaller $(i,j)$.
This is a signature that we have a fairly abrupt partition between the wealthiest and the mass (a mass that includes the $10\--100$ ``affluent'' wealthiest out of $N=3\,600$), the former influencing the overall fate by attracting wealth from all other agents~\cite{Bouch00, Piane, Heinsa, Chakr15, Alvar}.

Thus, our indicator tells how to define two ``effective classes'', separated by the ``water divide'' of wealth flow, on the average. The fact that it is not a single agent is of good omen for the application of the model. Playing with the basic GBM parameters $(N,\beta$ and the ratio $w_p/w_1)$, the relative size of this class would evolve from this $\sim 0.1\%$
 to another fraction. The inner fluxes inside each class could be studied in order to further partition each of these ``basin'' of the water divide picture. Then, a reverse procedure could help establishing sensible clever GBM nonlinearities, for example a  $\beta(N,w)$ dependence, that could help mapping actual econometric data sets into GBM models with somehow socially agnostic assumptions. The dynamics of each subset equally deserves attention~\cite{Adamou, Gabai16, Berman, Liu}.

What we can do with modest effort is to examine Fig.\ref{fig:five}(b) in more detail to get qualitative clues on the dynamics. If we look at the lower part of the distribution, well below the ``water divide'', the large intermittency features are more and more quickly washed out as we go to small wealth,  and the main governing factor seems to be the negative impact of the aggregate wealth. If we now look at the population across all times, in Fig.\ref{fig:six} color map, we see that, as is logical, the indicator tends to vanish for the least wealthy agents (blue-green shades at the bottom right). The absence of randomness of the indicator’s sign occuring for the smallest wealth suggests that their inter-exchange (that takes place in principle through the forced averaging process which includes their own collective fluctuation) is a minority mechanism. Their fate is dominated, as is obvious from the overall distribution in time, by the influences of wealthiest: the trickle-down effect in recovery phases (when the wealthiest give or ``emit'' wealth) or the austerity effect in collapse phases. 

We have now examined a canonical version of our multiplicative wealth model. Once sufficiently large wealth are generated, there is a regime of boom and busts, with abrupt collapses and slower revivals, due to the coupling induced by the constant average. Fluctuations of the richest are enough to cause large swaths of the population to be affected within short times. More analysis would entail tools such as momenta or Laplace transforms, with a scope of finding ``excited modes'' of the distribution. While this has a simple sense around equilibrium, we have no clues as to what are excited modes in a deeply nonstationary and broken ergodicity context. However, if we tame the nonstationarity, we may recover a system amenable to an eigenmode (fundamental and excited modes) analysis. The idea would then be to track how these excited modes behave when reintroducing nonstationarity. Taming nonstationarity and inequality intermittency is just the purpose of the following.

It is tempting to think of ``nudging'' the underlying laws so that this intermittency regime and its induced collapses are avoided. This entails avoiding the advent of large wealth if we want to maintain that $\beta$ itself represents a ``psychological constant'', the amount associated to risk at daily scale (scaling with the square root of time, basically). Forbidding large bets to owners of large wealth per se, even when they are designed well within regulatory barriers, would be a too directive way to interfere with the economic microscopic decisions. So in Sec.IV below, we introduce a modification of the basic probability law of our multiplicative process, $\Pi(\lambda)$, that averts the build-up of ``extremely extreme'' wealth. Stationarity is viewed as the obtainment of a fundamental mode of the system, with the corollary that obtainment of excited modes in the same frame is natural, but we shall not study their dynamics in the present work.

\section{IV: Agent ensemble with reset average and weakly wealth-dependent multiplicative process}

To avoid the advent of large wealth, we first define a ``status indicator'', based on wealth here. The role of ``status'' as a general factor in setting the price of exchanges dates back to Aristotle and was revived in social science and anthropology by P. Jorion~\cite{Gilen, Vital, Jorio}, with the aim of escaping the conventional wisdom of prices fluctuating around a ``fundamental price'' that an undistorted market is supposed to reveal (see also the final discussion). Aside such general views, a ``status indicator'' would be a good channel to link in the future our deliberately limited study to more complex ones with a developed social account~\cite{Levy, Alvar, Galle, Chatt16}. We cannot use the ``intrinsic'' classes defined by the ``water divide'', because they correspond to time average of a nonstationary process, so that they are not known until the relevant fluctuations did take place. We thus find it sensible to define a continuous status without connection to the nonstationary dynamics. 
Here, our status indicator denoted $S_j$ is a simple homographic function based on the comparison of wealth above poverty to average wealth:

\begin{equation}
S_j(t)=\frac{w_j(t)-w_p}{w_1+\left(w_j(t)-w_p\right)}
\label{status}
\end{equation}
So it tends to unity for large wealth, is one-half for $w=w_1+w_p$  ( $w =1\,400$ in our case), and tends to zero for $w \rightarrow w_p$ . We then modify $\Pi(\lambda)$ to introduce a counter-acting bias. Specifically, we make $\Pi(\lambda)$ status-dependent ($S$-dependent) using a skew factor  $\varepsilon$  as follows:

\begin{eqnarray}
\Pi_{\varepsilon}(\lambda)&=&\Pi_1(\varepsilon,S)\;\textrm{rect}\left(\frac{\lambda-(1+\varepsilon S)}{2\beta(1+\varepsilon S)}\right) \nonumber \\
&=&\frac{1}{2\beta(1+\varepsilon S)}\;\textrm{rect}\left(\frac{\lambda-(1+\varepsilon S)}{2\beta(1+\varepsilon S)}\right)
\label{statusPi}
\end{eqnarray}

So $\Pi_{\varepsilon}(\lambda)$ is centered at its centroid $\bar{\lambda}=1+\varepsilon S$ , spanning uniformly the range $1-\beta \rightarrow 1+\beta(1+2\varepsilon S)$. Technically, we obtain it as $(1+\beta[1-2\,\textrm{rand}])(1+\varepsilon S)$ instead of $1+\beta[1-2\,\textrm{rand}]$ for Eq.\ref{eqnw1wplambda_3}(c), rand being the usual uniform random variable in $[0,1]$.

We see that $\varepsilon$ plays the role of a wealth amplifier if $\varepsilon>0$: the wealthiest entities turn exchanges to their advantage, a well-known fact, evidenced by Piketty using the yields of the funds of US universities~\cite{Piket}. It plays the role of a taxation mechanism if $\varepsilon<0$, pushing the average factor $\bar{\lambda}$ to more than unity for the poorest and less than unity for the wealthiest.

We shall see that, surprisingly, very small skew factors $\varepsilon$ are sufficient to avoid the build up of large wealth. Or not so surprisingly, as it boils down to inhibiting the growth of the very few large wealth agents of the histogram that was built up in years, not days. If this distribution is diminished at large wealth by a modest factor per octave, then, since concerned agents are 10 or more octaves wealthier than the median  ($2^{10} \sim 1\,000$, cf. Eq.\ref{xnkmax} and the $10^{2.74}$ factor that gave the maximum N-ile at $550(w_1-w_p) \simeq 3.30\,10^5)$, it provides a significant inhibition of the tail. We will report below efforts to quantify the stationary distribution that results from this modification. Let us examine the impact of $\varepsilon\neq0$ through simulations first, using all other parameters as before. 

\begin{figure}
   \includegraphics[width=\columnwidth]{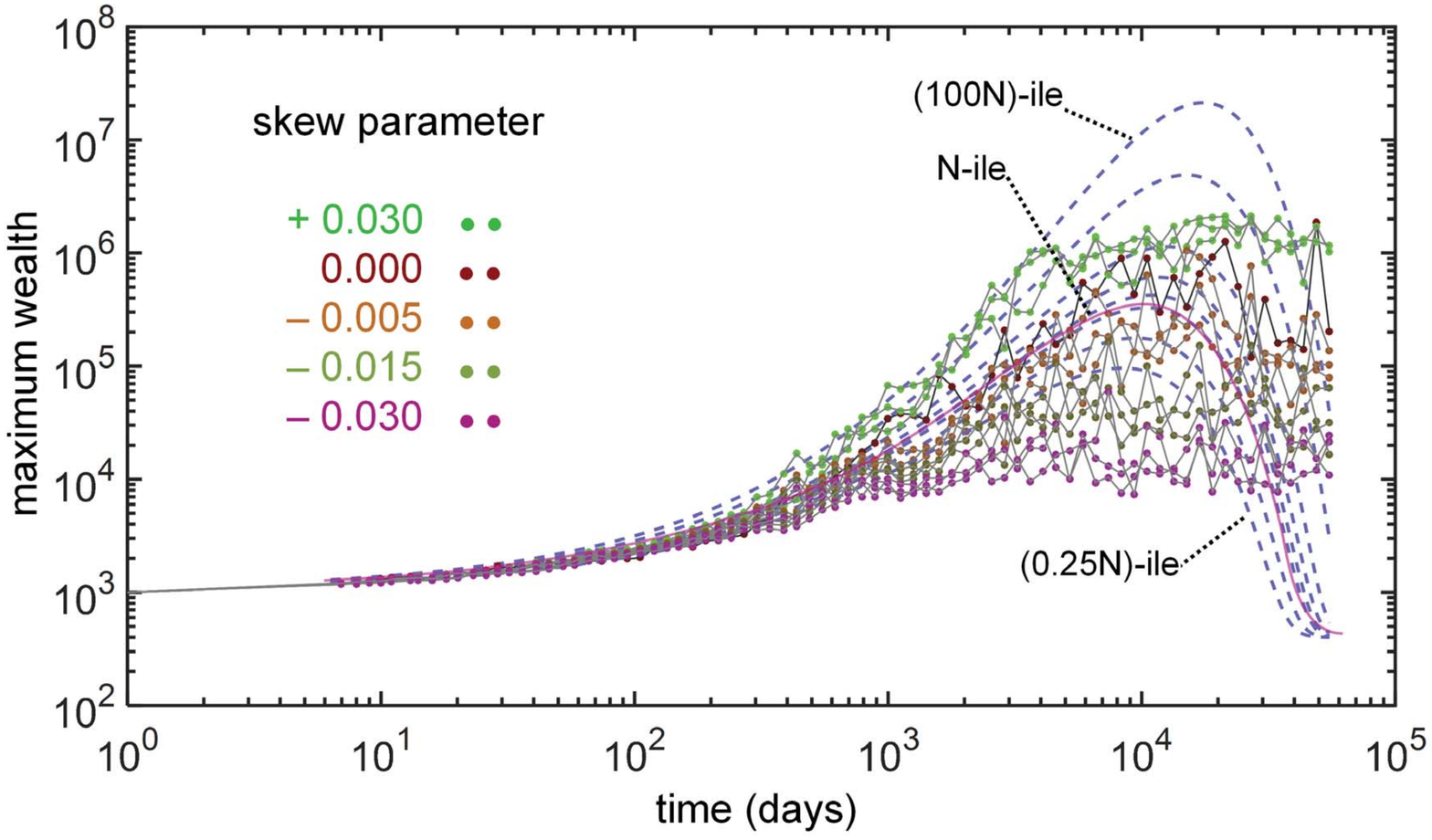}
  	\caption{(Color online): Fate of the richest wealth under different assumptions for the skew parameter $\varepsilon$. The analytical curves provided for the simple case of no skew and no reset in Sec. II are left as a guide to the eye.} 
  	\label{fig:seven}
\end{figure}

In Fig.\ref{fig:seven}, we examine the wealthiest agent evolution for various skew parameters $\varepsilon$. We have left as a guide the curves of the $(N/k)$-iles of Sec.II. We see that the modification does have the expected effect, and that this effect is large even for $\varepsilon$ values as small as $\varepsilon=-0.005$. The effect is not symmetric as we already evolve at $\varepsilon=0$ in a situation of extreme wealth reaching a large fraction of the total wealth: there is little room to expand more the wealth, and the distribution clearly saturates (akin to wealth condensation) for the $\varepsilon=0.03$ skew (positive feedback) parameter shown here. For negative skew parameters $\varepsilon$, we see that the wealthiest values clearly diminish by about 1.5 decade for $\varepsilon = -0.03$. At the same time, relative fluctuations tend to diminish (graphically obvious in log scale).

\begin{figure}
   \includegraphics[width=\columnwidth]{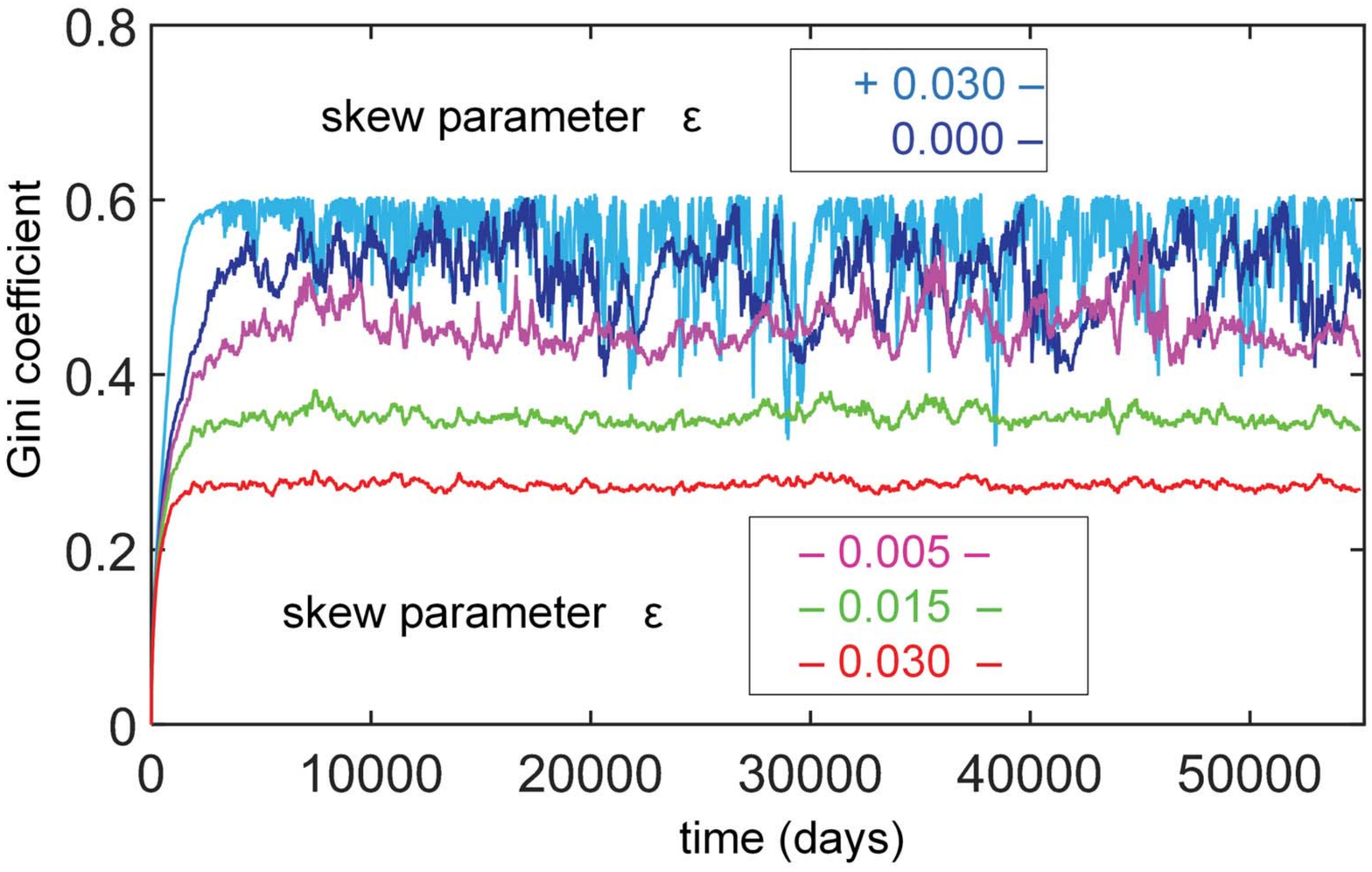}
  	\caption{(Color online): Gini coefficient under different assumptions for the skew parameter $\varepsilon$ as indicated.} 
  	\label{fig:eight}
\end{figure}

\begin{figure*}
	\includegraphics[width=13.5cm]{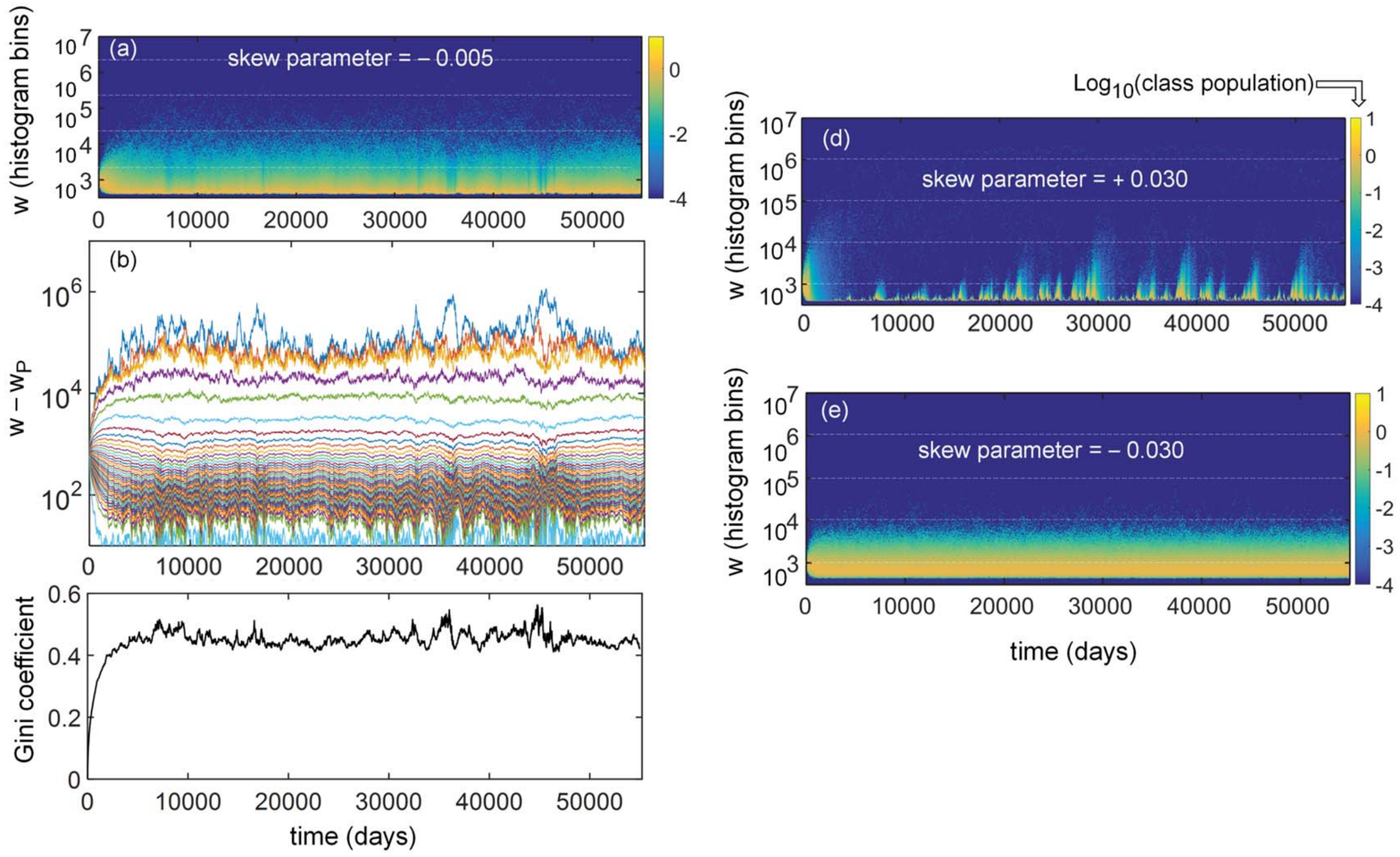}
  	\caption{(Color online)(a-e) For a simulation with a small negative skew parameter $\varepsilon=-0.005$, the evolution is depicted: (a) the histogram of wealth distributions as a color map coded as (d-e); (b) selected sorted wealth from highest to lowest, showing the few occurrences of anti-correlation at times $36\,000$ and $45\,000$ ; (c) the Gini coefficient, with spikes at those moments; (d) and shows the histogram for a case of large positive skew parameter $+0.03$, with extreme intermittency; (e) shows the case of a large negative skew parameter $-0.03$, with a very steady situation.} 
  	\label{fig:nine}
\end{figure*}
In Fig.\ref{fig:eight}, we show the Gini coefficient evolution for various values of the skew parameter $\varepsilon$.  The largest fluctuations are those of $\varepsilon = 0.03$, but, as said, they ``bump'' on the ceiling of wealth saturation. Also, due to the positive feedback and the subsequent roll-off of large wealth, variations are very quick. For negative values of $\varepsilon$, now, fluctuations of the Gini coefficient diminish clearly even for $\varepsilon = -0.005$. And as the distribution finds a stationary shape, these fluctuations nearly vanish for $\varepsilon = -0.03$.

In Fig.\ref{fig:nine}, we illustrate the evolution for  $\varepsilon = -0.005$, an interesting limit in our simulations window (a window intended to describe boom-and-bust cycles at the $\sim 2$ centuries scale). We see in Fig.\ref{fig:nine}(a) that the distribution histogram is stable most of the time, but still vulnerable to intermittent limited collapses, at $t=36\,000$ and $t=45\,000$. At these points, the wealthiest agent clearly exhibits an anti-correlation with most others, and shakes the whole distribution, see Fig.\ref{fig:nine}(b), as identified on similar earlier graphs. These are the points of surging Gini coefficient, as is seen on Fig.\ref{fig:nine}(c), albeit by a moderate amount. For the sake of comparison, we provide on Fig.\ref{fig:nine}(d,e) histograms of the two extreme and contrasted situations $\varepsilon = \pm 0.03$. In the positive case, we see that the evolution is a tale of few moments of ``shared prosperity'' and many moments of utter inequality. But as soon as an agent takes over, it acts over the whole distribution, and it is ``wagging'' all the distribution very soon ($\sim 1000$ days scale, that corresponds to periods such as revolutions). And in the negative case, stabilization is obtained early, only gentle fluctuations are seen in those slices of rarefied statistics, indicating by contrast that all slices below operate in a stationary stabilized and balanced regime.

\begin{figure}
   \includegraphics[width=\columnwidth]{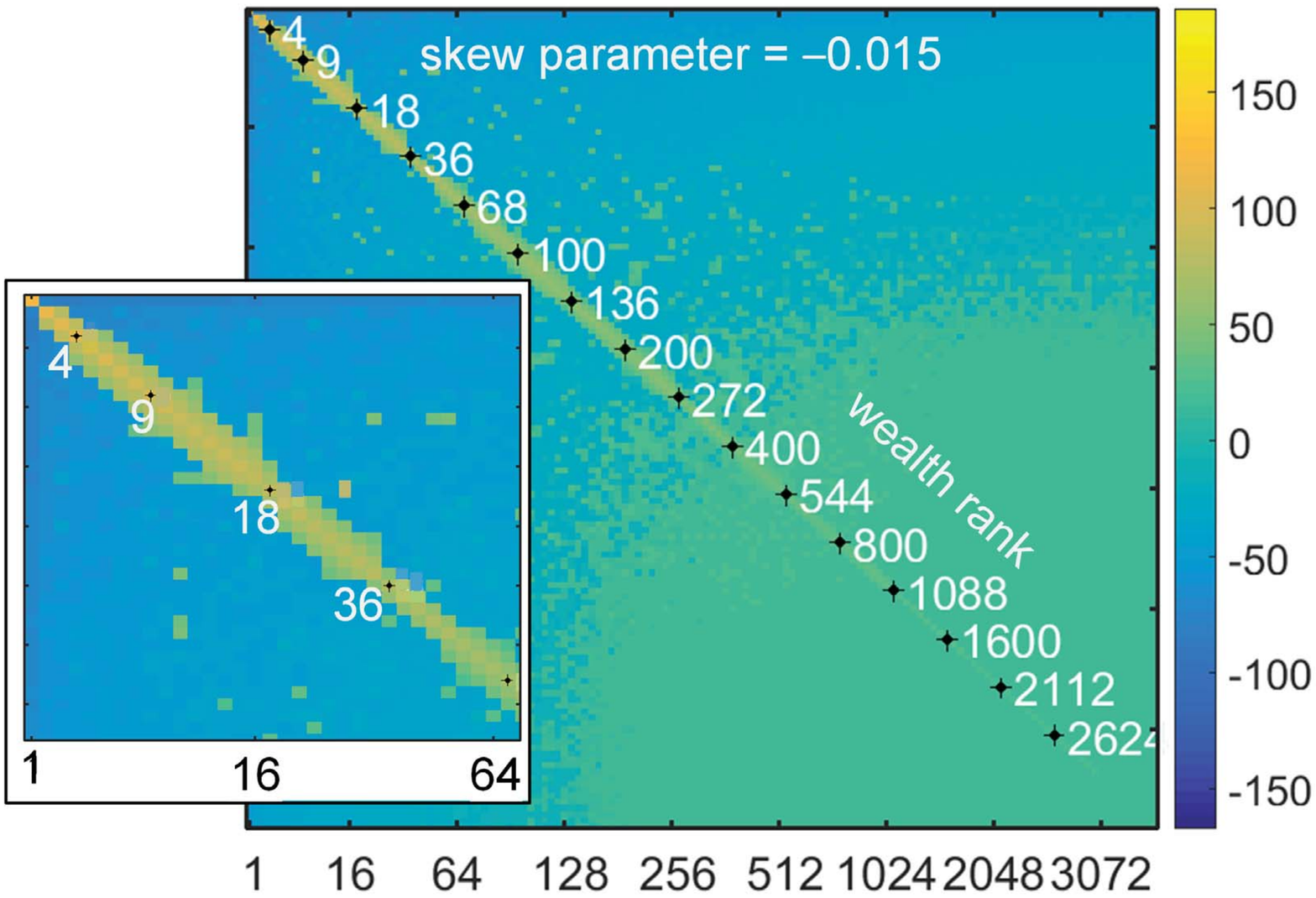}
  	\caption{(Color online) Map of correlation of wealth variations, $C_{i,j}$ , among sorted agent records, for a negative skew coefficient $–0.015$ as in Fig.\ref{fig:six}. The transition from positive to negative fluxes is smooth and lies around $i,j\sim 300$, it is smooth and has no strong values, except close to the diagonal (artifact of correlation when ranked agents ``cross'').} 
  	\label{fig:ten}
\end{figure}

In Fig.\ref{fig:ten}, we show the same correlation map with the quantity $C_{i,j}$ defined above as in Fig.\ref{fig:six}, but for a negative intermediate value of the skew parameter $\varepsilon = -0.015$. The ``water divide''  line between the richest and the poorest is now lying around a gentle intermediate value, around the 300-th of the ranked agents. This is likely to coincide with the mean wealth $w_1$, but we have not examined this in detail. There are a few spurious positive correlations on lines next to the diagonal. We believe they stem from the choice of using an indicator based on sorted distribution. The sorting introduces correlation between adjacent ranks, if they both mostly suffer from other agents, but just spend some time ``crisscrossing''. Anyway, it does not perturb much the overall picture, but rather indicates that the next steps in such simulations would be to understand, as in many current physics problems, the role of correlations~\cite{Heinsa, Chakr15, Galle, Chatt16}, which is a general concern in the newly emerged considerations of GBMs~\cite{Gabai16, Liu, Berman, Adamou}.

\begin{figure}
   \includegraphics[width=\columnwidth]{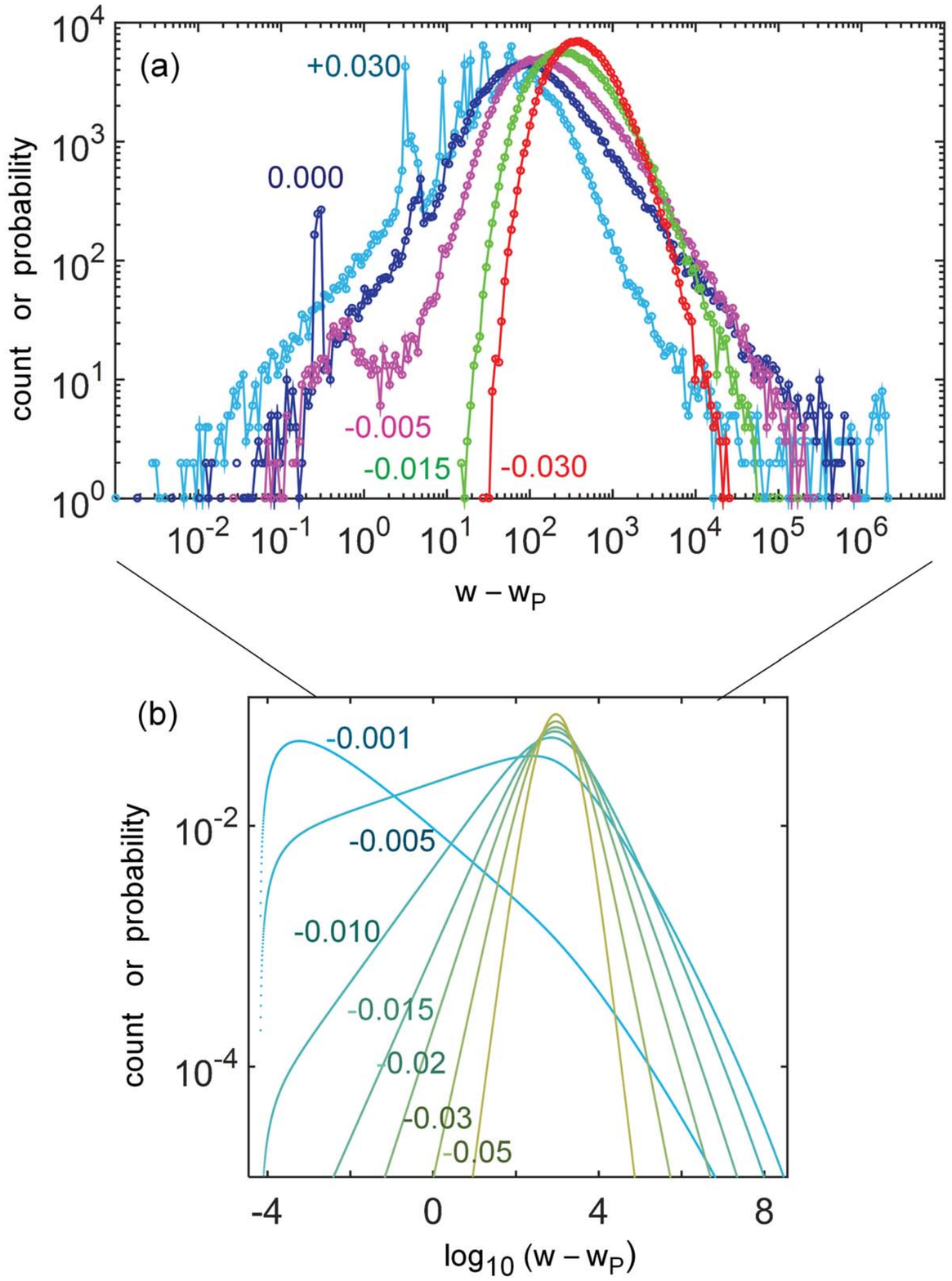}
  	\caption{(Color online)(a) Histograms of ($w-w_p$) on log-log scale, for various skew parameters as indicated at a large enough time to escape the initial phase. Note the spikes on the left side of the distributions down to parameter $-0.005$, that are created at the ``bust'' events of wealthiest agents. Spikes at the right end are the few wealthiest agents; (b) Eigenmodes associated to the largest (unity) eigenvalue for the continuous model without the normalization effect, for negative values of the skew parameter. For too weak values, there is no solution within the simulation space, the distribution is at the verge of unstable negative drift, and hence the eigenmodes explore boundaries.} 
  	\label{fig:eleven}
\end{figure}

In Fig.\ref{fig:eleven}(a) we show the histogram of wealth distribution at $t=55\,000$ for various simulations. Depending on the skew parameter $\varepsilon$, we see clear indices of the mechanisms operating for these different distributions. For a skew parameter $\varepsilon \geq 0$, we see strongly populated peaks that ``scar'' the left side of the distribution, some of them not so far from the main peak. The distribution is broad, and we also see on the large wealth tail a few peaks with one or a few individuals that are above the trend of the tail at lower values. Both indications are logical with the mechanism of ``the tail wagging the dog''. We see now somewhat more in detail that there are highly populated sets of agents that were in some narrow interval close to $w_p$, and that benefited from an upward kick when the wealthiest agents fluctuated downward. The statistical characteristics of these intermittent bunches may be an interesting part of future work, in relation with GBM dynamics.

As for the distributions for $\varepsilon<0$, they clearly get narrower as $\varepsilon$  becomes more negative, and logically, they tend to become stationary. The equilibration time~\cite{Patri,Berman,Gabai16} is shorter for the more negative values of  $\varepsilon$. We now have a clearer view of how the distribution is curtailed on the high end: by a decade or so around $w=10^5$, for the case $\varepsilon=-0.015$ \textit{vs.} the reference $\varepsilon$ =0.
The fact that such a small skew could avert the large fluctuations initially surprised us (somehow as the diverging feedback of inequality in the reallocation+GBM model of Ref.~\cite{Berman} surprised their authors). But considering the effects of the residual drift that we explored in Sec.II, it is not so surprising that a very limited but ``daily'' drift acts in such a large manner.
 
Since a stationary distribution results, we can determine it, thus allowing comparison with the broad literature addressing this topic ~\cite{Chatt04, Chatt07,Yakov, Heinsa, Saich08,Chatt16,Chakr09,Gabai16,Liu}. It is possible because we can assume that the normalization mechanism does not play a role anymore ~\cite{Gabai99,Saich08}. The signature of this mechanism was the set of peaks or bunches on the left side of the distributions. They are still seen in the limit case $\varepsilon = -0.005$ and are associated to very modest wealth steps values (less than unity, hence $\sim (w_1-w_p)/600$), thus to a small impact of the normalization acting on modestly large wealth. Such effects apparently rarefy and vanish for more negative values within our simulation bounds.

To find the stationary limit, we do not need detailed balance as exchanges are not explicitly accounted in our model. We have to solve the functional integral equation:

\begin{equation}
P(w)=\int_{w_p}^{+\infty}{P(w')} \Pi_{\varepsilon}(\lambda)\,d\lambda,
\label{PPiepsilon}
\end{equation}
which accounts for a single time step in a mean-field view, assuming as a boundary condition that no probability flux comes from the region close to $w_p$. We should also care that the probability $\Pi_{\varepsilon}(\lambda)$ is actually coupled to $w$ by the status factor. A full notation would be $\Pi(w',\lambda)$ or $\Pi(S(w'),\lambda)$. However the numerical values in this equation are spread on decades. So we can transpose this in the $\{x,\ell\}$ space, whose (now stationary) probability law and multiplier law are  respectively $\hat{P}(x)$ and $\pi(\ell)$. With the now additive algebra, given that $x'+\ell=x$, and with an adequate redefinition of $\pi(\ell)$ into $\pi_{\varepsilon}(\ell)$ to incorporate the status $S$, we find:

\begin{equation}
\hat{P}(x)=\int_{-\infty}^{+\infty}{\hat{P}(x-\ell)} \pi_{\varepsilon}(\ell)\,d\ell
\label{PPiepsilonx}
\end{equation}
This form is reminiscent of a convolution (as it should).  But due to the dependence of $\pi_\varepsilon(\ell)$  on $x$  through the status $S$, it is not a convolution. Nevertheless, it is possible to solve for this equation in the form of an eigenvalue problem onto a discretized and uniform set $\{x_m=m\,\Delta x  \}$ of  $M$ values of $x$. Also, in the limit of a large enough number of iterations, we can remember that the fate of the distribution was given at $\varepsilon=0$ by its drift and second momentum (diffusion constant, essentially $\beta$ for us)~\cite{Sorne97,Sorne98,Yakov,Saich09}. While we do not have a theorem to extend this to the case $\varepsilon \neq 0$, we conjecture that the first-order findings we want can be made retaining this assumption. So the matrix representing Eq.\ref{PPiepsilon} as an operator on $\hat{P}$ is built up as a Toeplitz matrix, with the $n$-th diagonal having a coefficient $\pi_\varepsilon(n\,\delta x)$ . This introduces a constraint as we want to span several decades (12 decades, see below) , so that even in a large ensemble  $\{x_m=m\,\Delta x  \}$ with an $M$ value of a few thousand, only a few of these values fall within the modest range located between extrema of $\ell$ , essentially $\log(1\pm\beta)$, apart from the small corrective action that describes the law $\pi_{\varepsilon}(\ell)$. However, with some care on this sampling, and given the assumption mentioned above, we could find significant solutions with matrix sizes $M$ of a few thousand, and typically $10\--20$ filled diagonals. Then, a significant $\pi_{\varepsilon}(\ell)$ can still be put up by operating on the rows of the matrix, shifting the centroid of the $\pi_{\varepsilon}(\ell)$ distribution to the proper drift-induced value equivalent to Eq.\ref{statusPi}, but not taking into account the modified width of the distribution. This introduces some second-order effects when it comes to converge to a stationary distribution.

Fig.\ref{fig:eleven}(b) shows the result of this approach using a set spanning 12 decades. We indeed find that the largest eigenvalue of our matrix is nearly unity, and its eigenvector is generally a bell-shaped distribution. For too weak negative values of  $\varepsilon$, about $|\varepsilon| < 0.002$, the distribution peaks near the truncating boundary condition that we implemented at the lowest $x$ values. Just because we do not normalize the rows of the matrix at its ``corners'', the probability can be ``dissipated'' there, and the drift+diffusion processes must accommodate this numerical boundary. Such a typical appearance, provided here for $\varepsilon=-0.001$,  means physically that the distribution is still dominated by the negative drift and is not stationary, essentially as in Sec.II (but not really as in Sec.III as Eq.\ref{PPiepsilon} ignores the total wealth renormalization).

For more negative values of $\varepsilon$ , we find stationary bell-shaped curves. They present most of the characters of the actual distribution: there is notably a shift to the right and to a narrower distribution that is quite comparable to the simulation of Fig.\ref{fig:eleven}(a). However, we suspect that our solving procedure is not accurate enough to attempt a meaningful fit: a better numerical solution should be sought. Even without an exact account of our model, we can nevertheless discuss its benefits. 

An interesting exercise around the class of stationary distributions that are currently under scrutiny is to attempt to look at the relaxation rates of the excited states~\cite{Berman}. Very plausibly, the relaxation rates will be faster for the higher excited states, and the first excited state ~\cite{Gabai16} gives a measure of the most relevant time scale for external shocks that affect inequality, i.e., that affect differentially the rich and poor agents. Nonlinearities could also be investigated in terms of the effect of correlation. Naively, letting an initial state evolve from the two linear combinations $|p\rangle\pm|q\rangle$ of the $p$-th and $q$-th modes could modify the relaxation rate due to the status term, as this latter is not respecting any of the mode orthogonality conditions.

\section{V. Discussion}
The quest for econophysics models to understand inequality recently evolved from the study of stationary distributions~\cite{Sorne98, Levy, Piane, Chakr15, Saich08, Malev, Alvar, Piket,Galle, Chatt16, Vital} to the much more fascinating issue of the distribution dynamics~\cite{Gabai16, Berman, Adamou, Liu}. The simple tool of GBM gives insight to such models, not least because it addresses some limitation of mainstream economics (e.g. a bounded utility function), but also because it helps tackling the false intuitions that arise when ergodicity breaking is not properly taken into account~\cite{PRL, Gamble}.

Setting up a model with a finite number of agents and a fine-grain (``daily'') time discretization, we have introduced a nonstationary regime of sustained intermittency by using the normalization of the total wealth. A lively dynamics emerges, with still much to analyze. The number of agent used in the simulations ($N=3\,600$) was enough to attain in a reasonable time scale the situation of an extreme degree in wealth capture by a few individuals, whose decisions then impact the fate of all agents within short times (say, few months), but not in any circumstances, rather only once a crisis is triggered~\cite{Saich09}. The way the fluxes have their signs undergoing inversion when scanning in a sorted agent distribution (Fig.\ref{fig:six} and Fig.\ref{fig:ten}) is one of the most meaningful signatures we got. It defines two natural subsets across a ``water divide'' of wealth flow. Their coupled fate can then be captured in a nutshell by the ``tail wagging the dog'' metaphor. It could be applied to actual statistics or to any of the many more explicit models. Sticking to the dynamics of subsets and aggregates, we would have tools that remain intrinsic or ``agnostic'' enough along this line. We are also aware that calibration of GBMs is in infancy and will by itself reveal several features of interest or even prompt new uses of GBMs.

Also, generally speaking, when inside a general large system, a subsystem presents sufficient stationarity, its degree of redistribution could be studied with an appropriate scaling of $N$ ~\cite{Chatt04, Patri, Yakov, Chakr15,Saich09, Malev}(we considered only GBMs coupled by the normalization, but an absence of reallocation in the sense of ~\cite{Berman}). The comparison is not limited to wealth as traditionally quantized in economics, it can be adapted to various cases around the general balance idea.  This principle entails the statistical fairness in the microscopic trend (as much chance to get more than less in a single event), but nevertheless the small gain of a minority in relative terms appears to be self-amplifying, even weakly. This could be for instance the fate of fashionable topics in a domain of science, where the developing trend first looks like a fair reward, but if it becomes a dominant trend, it can lead to too many followers and production of apparent knowledge with little actual relevance in a majority of cases~\cite{Chatt16b}. As several of these domains are not as long-lived as human economies, it makes sense to start simulation from an apparently stationary distribution. Then, the issue of dynamics has a simple focal point: how much the first main crisis can be anticipated (see the rich studies on firm births and deaths~\cite{Saich09}), or more precisely: can we find how the onset of crisis can be described in a more detailed fashion, e.g. by connecting its probability to all average features/momenta of the wealth distribution or of its underlying log-scale counterpart ?

In the economic domain, the large intermittency is often linked to the Black Swan paradigm of Nassim. N. Taleb, which relates rather to bust phases, but it is more difficult to assign ``white'' or``withish'' swans in the positive boom phase of cycles among the impact of technological, societal or political changes. Microscopic studies of different sectors and their interactions could benefit from the a comparison with our kind of stochastic model in this respect.   

After such ``microscopic'' considerations, let us take briefly a broader perspective: Our initial intention, that we hope to be still present in the result, was to put Piketty's historic-economic narrative~\cite{Alvar, Piket} and some of its ``obvious'' consequences into a model that would go one step beyond the stage of ``riches become richer'', with the further prospect of shedding light on how the underlying networks and their concentration effects operate in terms of statistical distribution of economic and social variables~\cite{Galle, Vital, Jorio}. On this way, we were faced with the fact that growing inequalities and nonstationary distributions occur even in the simple paradigm of apparently local fair exchange, a result that stems from the natural drift of a zero-sum-exchange in log-scale terms. There was not much appearance of such issues in the econophysics literature until a few years ago. Then, as the issue of evolving inequalities became more paradigmatic with data available across the 2008 crisis, the dynamics of inequality and the capability of GBMs to describe them became a center of intense attention~\cite{PRL, Gamble, Adamou, Gabai16, Berman, Liu}. Indeed, there is a interesting parallel between on the one hand Piketty's ``divergence'' of the $r > g$ picture, whose meaning is rather historical than a precise econometric exercise (hence Piketty stops short of a divergence model), and on the other hand the nonstationarity and ergodicity breaking of GBM ensembles~\cite{PRL, Gamble, Adamou}.

Taming this nonstationarity involves no less than assessing whether our economies run in a near-equilibrium fashion, or more deeply out-of-equilibrium~\cite{Berman} even though sociology and economics can track a number of slowly drifting items that are deceivingly suggestive of an adiabatic evolution restoring an equilibrium induced by the noise of external shocks.

 In our case, instead of introducing an explicit taxation mechanism having an extra variable and entailing no less than the prerogatives of a ``State'' to run it,  we chose a more implicit or self-contained approach: Our above presentation of the equilibrating mechanism as relying on ``status'' is drawn from anthropological considerations brought into the realm of finance and economics by Paul Jorion, who found that the law of supply and demand was only marginally verified in actual communities submitted to extreme risks of subsistence (fisher communities for instance). Rather, based on Aristotle inspiration (picked up from Karl Polanyi's writings), the survival of the community, and the reproduction of the member's status throughout its social and economic exchanges, was felt to be a more general factor, preventing prices to fall too low, or on the contrary, causing the buyers to have the last word even in periods of high demand as part of their accepted higher status.

We have therefore introduced an equalizing mechanism akin to taxation directly as an exchange bias that can be seen also as an average price bias. The fact that with enough corrective strength, the distribution turns from nonstationary to stationary is no surprise. The interesting point that would not have been guessed easily at first is that a quite limited skew or bias, on the order of 1\% in the daily transaction, is sufficient to strongly suppress the advent of inequality-induced crisis. It is admittedly not obvious to connect (and possibly contrast) a small daily bias on the one hand and, on the other hand, the current conventional wisdom that yearly tax rates for the affluent must lie somewhere in a range of 15-60\%. At a time when economic models are under criticisms from several points of views~\cite{Alvar, Galle, Chatt16}, we believe that the knowledge brought by our simple model is a source of inspiration for all three communities of physics, econonophysics, and the broader social sciences that embed economics. This inspiration rests on a fertile ground thanks to the recent consideration of all GBM properties and of their subtle consequences in inequality models~\cite{PRL,Liu,Gamble,Adamou}.

\medskip

\section*{ Acknowledgments}
Henri Benisty thanks Paul Jorion and Serge Boucher for useful hints and presentation remarks about the model. He is also extremely thankful to Didier Sornette for advices on the use and existing knowledge of multiplicative processes in econophysics. He finally thanks reviewer A for indicating timely studies of the GBM. 


\end{document}